\documentclass{aastex61}

\newcommand{\co}{CO$_2$}
\newcommand{\fice}{F_{\rm ice}}

\newcommand{\lsun}{L_\odot}

\newcommand{\tsf}{T_{\rm sf}}

\received{}
\revised{}
\accepted{\today}
\submitjournal{ApJ}

\shorttitle{Seafloor Weathering and Planetary Habitability}
\shortauthors{Chambers}

\begin{document}

\title{The Effect of Seafloor Weathering on Planetary Habitability}

\correspondingauthor{John Chambers}
\email{jchambers@carnegiescience.edu}

\author{John Chambers}
\affil{Carnegie Institution for Science \\
Department of Terrestrial Magnetism, \\
5241 Broad Branch Road, NW, \\
Washington, DC 20015, USA}

%
%
\begin{abstract}
Conventionally, a habitable planet is one that can support liquid water on its surface. Habitability depends on temperature, which is set by insolation and the greenhouse effect, due mainly to \co\ and water vapor. The \co\ level is increased by volcanic outgassing, and decreased by continental and seafloor weathering. Here, I examine the climate evolution of Earth-like planets using a globally averaged climate model that includes both weathering types. Climate is sensitive to the relative contributions of continental and seafloor weathering, even when the total weathering rate is fixed. Climate also depends strongly on the dependence of seafloor weathering on \co\ partial pressure. Both these factors are uncertain. Earth-like planets have two equilibrium climate states: (i) an ice-free state where outgassing is balanced by both weathering types, and (ii) an ice-covered state where outgassing is balanced by seafloor weathering alone. The second of these has not been explored in detail before. For some planets, neither state exists, and the climate cycles between ice-covered and ice-free states. For some other planets, both equilibria exist, and the climate depends on the initial conditions. Insolation increases over time due to stellar evolution, so a planet usually encounters the ice-covered equilibrium first. Such a planet will remain ice-covered, even if the ice-free state appears subsequently, unless the climate receives a large perturbation. The ice-covered equilibrium state covers a large fraction of phase space for Earth-like planets. Many planets conventionally assigned to a star's habitable zone may be rendered uninhabitable as a result.

\end{abstract}

%
%
\section{Introduction}
Discussion of planetary habitability often begins with the concept of a star's habitable zone \citep{kasting:1993}. Conventionally, this is defined as the region around a star where a rocky planet can have liquid water at its surface. Potentially habitable planets are assumed to have modest reservoirs of water and carbon dioxide, some contained in the atmosphere, and the rest condensed on the surface or buried in the planet's crust and mantle. The surface temperature is set by a combination of the insolation and the atmospheric greenhouse effect due to \co\ and water vapor. Planets in the habitable zone have the right combination of atmospheric composition and insolation to allow liquid water on the surface.

The habitable zone is a useful starting point, but it has some limitations. One drawback is that liquid water is not the only prerequisite for life to exist. Life, especially advanced life, probably requires a number of other conditions to exist before it can form and thrive. Two factors that are relevant in this study are the surface temperature of a planet and the partial pressure of \co\ in the atmosphere.

Complex, multi-celled organisms probably require a fairly narrow range of temperatures in order to survive and reproduce. \citet{silva:2017} present a number of arguments showing that advanced animal life on Earth, or its evolutionary precursors, needs temperatures in the range $273\le T\le 323$ K in order to operate. This is much narrower than the temperature range of liquid water, which extends up to the critical temperature 647 K. Complex life on Earth relies directly or indirectly on oxygenic photosynthesis. Most known photosynthesizers are unable to operate when the partial pressure of \co\ falls below about $10^{-5}$ bar \citep{caldeira:1992b}. There may also be an upper limit on the \co\ level that advanced life can tolerate, either directly or through its effect on ocean pH levels \citep{schwieterman:2019}.

A second limitation of the habitable zone concept is that it is an instantaneous measure of the state of a planet's climate. It does not take into account the increase in insolation over time as the planet's star grows more luminous. The usual definition of the habitable zone also doesn't consider changes in a planet's geological activity that can affect its atmospheric composition.

The climate on geologically active planets like Earth is thought to be stabilized by temperature-dependent weathering. Weathering involves  reactions between the planetary surface and the atmosphere that alter the atmospheric pressure of \co\ and the strength of the greenhouse effect \citep{walker:1981}. Weathering of continental rocks removes \co\ from the atmosphere/ocean system. Carbon dioxide is replenished by volcanic outgassing from midocean ridges and island arcs. Weathering rates increase with temperature, and land-based weathering essentially ceases below the freezing point of water \citep{walker:1981, marshall:1988, berner:1994, lenton:2018}. Outgassing rates are independent of temperature to first order. A high atmospheric \co\ pressure leads to a high surface temperature and rapid weathering, and vice versa. On geological timescales, this negative feedback can stabilize the \co\ level and keep the temperature in a fairly narrow range above freezing.

Models that estimate the size of the habitable zone usually assume that the weathering feedback operates and that the surface is ice-free as a result. However, the weathering feedback may not stabilize the climate if the outgassing rate is too low, or the  weathering rate is too high. The reason for this is that Earth-like planets can exist in more than one climate state. For insolations close to the modern value, Earth has two stable states: one covered in ice, and one largely ice-free \citep{budyko:1969, sellers:1969}. These two solutions exist because ice has a higher albedo at visible wavelengths than ice-free land or ocean, so the fraction of solar radiation that is absorbed is different in the two cases.

In some circumstances, a planet with a surface just above freezing can experience a weathering rate that exceeds the outgassing rate. In this case, the \co\ level will decrease, the temperature will fall, and the planet will freeze. Continental weathering now ceases, \co\ levels increase, and eventually the greenhouse effect becomes strong enough to melt the ice. Weathering then begins again. The timescale for changes in the \co\ level is much longer than the timescale for the surface to freeze or melt \citep{caldeira:1992a}. As a result, no equilibrium is established and the climate oscillates between glaciated and ice-free states \citep{mills:2011, menou:2015, abbot:2016}.

A further complication is seafloor weathering. To date, most studies of habitability have focused on continental weathering. However, weathering reactions also occur on the seafloor. These reactions remove \co\ dissolved in seawater, ultimately lowering atmospheric \co\ levels as the ocean absorbs excess \co\ from the atmosphere. Seafloor weathering, like continental weathering, depends on the local temperature, so both processes potentially provide a stabilizing feedback for the climate \citep{coogan:2015}. However, for seafloor weathering, the local temperature is not necessarily the same as the surface temperature because seafloor weathering takes place in rocks beneath the seafloor \citep{krissansen-totton:2017}.

Both types of weathering probably depend on the partial pressure of \co\ in the atmosphere, but not necessarily for the same reason. Land-based weathering rates depend directly on the \co\ level, modified to some extent by the effects of vegetation \citep{kump:2000, lenton:2001}. Seafloor weathering probably depends on the ocean's pH, which is influenced by atmospheric \co\ level \citep{krissansen-totton:2018}. However, seafloor weathering rates may be affected by other aspects of ocean chemistry as well \citep{coogan:2018}.

A central theme of this paper is that, while continental weathering ceases at temperatures below freezing, seafloor weathering may continue on an ice-covered world. This is plausible provided that the ocean doesn't freeze entirely. Complete freezing is unlikely to occur as long as heat is escaping from the planet's interior \citep{nakayama:2019}. Seafloor weathering will continue to affect planetary climate as long as there is adequate exchange of \co\ between the ocean and atmosphere. This is plausible because ice-free oases will be maintained around geothernally active regions, allowing \co\ to enter and leave the ocean \citep{lehir:2008}.

The continued operation of seafloor weathering during glaciated episodes can modify the climate evolution substantially. In this paper we explore climate and habitability on Earth-like planets subject to both continental and seafloor weathering, assuming that seafloor weathering operates even when land-based weathering stops. We also examine whether the climate is suitable for advanced life to exist according to the temperature and \co\ pressure constraints described above.

The rest of this paper is organized as follows. Section~2 describes the models used for the climate and weathering rates. In Section~3, we explore the climate evolution for a variety of situations, focussing in particular on the impact of seafloor weathering. The main results and implications are discussed in Section~4. Finally, Section~5 contains a summary.

%
%
\section{Model}
We consider the climate evolution of an Earth-like planet orbiting a Sun-like star. We assume that the planet undergoes crustal recycling due to plate tectonics or a similar process. The surface temperature is determined by a balance between stellar insolation $S$ and outgoing infrared radiation $I$, including the greenhouse effect due to carbon dioxide and water vapor. We assume that the water vapor content of the atmosphere adjusts instantaneously such that the troposphere is fully saturated \citep{kasting:1993}. The amount of carbon dioxide in the atmosphere increases due to outgassing from the interior, and decreases due to weathering reactions with rocks on land and the seafloor. 

The mean global surface temperature $T$ satisfies
\begin{equation}
\frac{S}{4}[1-\alpha(T, P)]=I(T, P)
\end{equation}
where $\alpha$ is the top-of-atmosphere albedo. To calculate $\alpha$, we use the formulae given by \cite{haqq-misra:2016} that are an empirical fit to results of a 1-D radiative-convective climate model. The formulae give $\alpha$ in terms of $T$, the partial pressure of \co\, denoted $P$, and the surface albedo $\alpha_S$. 

Following \cite{williams:1997}, we calculate the surface albedo, including the effect of clouds, assuming 50\% overall cloud cover, so that
\begin{equation}
\alpha_S=0.5\alpha_{LS}+0.5\alpha_C
\end{equation}
where $\alpha_{LS}$ and $\alpha_C$ are the albedos of the land/ocean and clouds respectively. The cloud albedo $\alpha_C$ is given by
\begin{equation}
\alpha_C=0.65Z-0.078=0.603
\end{equation} 
where $Z$ is the mean zenith angle of the Sun, taken to be $\cos^{-1}(0.5)$. 

For an ice-free surface with an Earth-like mixture of land and ocean, we use $\alpha_{LS}=0.0869$. This value reproduces the surface temperature of modern Earth for the modern isolation and \co\ level in our model. We assume that the albedos of ice-covered land and ocean are the same, and given by
\begin{equation}
\alpha_{\rm ice}=f_{\rm vis}\alpha_{\rm vis}
+(1-f_{\rm vis})\alpha_{nir}=0.656
\end{equation}
following \cite{kadoya:2019}, where $\alpha_{\rm vis}=0.8$ and $\alpha_{\rm nir}=0.5$ are the ice albedos in the visible and near infrared, and $f_{\rm vis}=0.52$ is the visible energy flux fraction for a Sun-like star.

We note that the global cloud cover may be reduced during episodes when the surface is ice covered. However, this will make little difference to the climate since the ice and cloud albedos are quite similar.

To calculate the outgoing infrared flux $I$, we use formulae developed by \citet{kadoya:2019} that give $I$ as a function of $P$ and $T$. These formulae are an empirical fit to a 1-D radiative-convective climate model with a fully water saturated troposphere generated by \citet{kopparapu:2013, kopparapu:2014}. 

The formulae for $I$ and $\alpha$ were obtained by fitting models with a limited range of conditions, in particular $10^{-5}\le P\le 10$ bar. At the lower end of this range, $I$ and $\alpha$ depend on $P$ in a simple manner, which suggests it is reasonable to extrapolate the formulae to lower values of $P$. We do this by assuming that $\alpha$ is independent of $P$ at \co\ partial pressures below 0.01 bar, and we perform a linear extrapolation in $\ln P$ for $I$ when $P<3\times 10^{-5}$ bar. As we will see later, habitable planets typically spend only a small fraction of their time with $P<10^{-5}$ bar, so this procedure is unlikely to affect the results substantially.

\begin{table}
\begin{center}
\begin{tabular}{c|c}
Coefficient & Value \\
\hline
$c_0$ &  7.1620e-1 \\
$c_1$ &  3.9036e-2 \\
$c_2$ &  1.3386e-2  \\
$c_3$ &  -2.4032e-3 \\
$c_4$ &   1.9475e-4 
\end{tabular}
\end{center}
\caption{Coefficients used in the formula for the stellar luminosity evolution.}
\end{table}

The insolation $S$ received by the planet depends on the mean orbital distance $a$:
\begin{equation}
S=S_0\left(\frac{L}{\lsun}\right)\left(\frac{a}{\rm AU}\right)^{-2}
\end{equation}
where $L$ is the stellar luminosity, $\lsun$ is the current value for the Sun, and $S_0=1360$ W/m$^2$. The luminosity increases over time as the star evolves. Here, we make a fourth-order polynomial fit to a stellar evolution model for a solar mass, solar metallicity star generated by \cite{bressan:2012}:
\begin{equation}
\frac{L}{\lsun}=c_0+c_1\left(\frac{t}{\rm Gy}\right)+c_2\left(\frac{t}{\rm Gy}\right)^2+c_3\left(\frac{t}{\rm Gy}\right)^3+c_4\left(\frac{t}{\rm Gy}\right)^4
\end{equation}
where $t$ is the age of the star beginning when the star reaches the main sequence, and the $c$ coefficients are given in Table~1. This expression is valid for the first 10.5 Gy of evolution.

We consider weathering that take place on land and the seafloor. In each case, weathering releases cations such as Ca$^{2+}$ into the ocean where  they react with dissolved \co, precipitating carbonates and thereby removing \co\ from the atmosphere/ocean system. We use the following expressions for land and seafloor weathering 
\begin{eqnarray}
W_L&=&W_{L0}(1-\fice)
\left(\frac{P}{P_0}\right)^{\beta_L}
\exp\left(\frac{T-T_0}{\theta_L}\right)
\nonumber \\
W_S&=&W_{S0} G
\left(\frac{P}{P_0}\right)^{\beta_S}
\exp\left(\frac{\tsf-T_0}{\theta_S}\right)
\label{eq-weathering}
\end{eqnarray}
where $T_0=288$ K and $P_0=3\times 10^{-4}$ bar, $G$ is the outgassing rate normalized to the outgassing rate on Earth today, and $\fice$ is the fraction of the surface covered in ice (assumed to be either 0 or 1). Also, $\tsf$ is the mean temperature in the solid layers beneath the ocean where seafloor weathering occurs. 

In the following sections, I refer to the constants $W_{L0}$ and $W_{S0}$ as the ``intrinsic'' land and seafloor weathering rates respectively. Thus, these represent the continental and seafloor weathering fluxes, relative to modern Earth, that a similar planet would experience when subject to modern Earth's temperature, \co\ partial pressure and seafloor spreading rate. The actual weathering rate is  likely to depend on a range of other factors such as the continental configuration, topographical relief, precipitation, chemistry of exposed rocks etc. These factors can change over time on a given planet, but we neglect these changes here.

In the model runs described below, we typically use $\theta_L=\theta_S=10$ K, which are comparable to values from other studies \citep{walker:1981, abbot:2012, coogan:2015, menou:2015}, although larger values of $\theta_L$ have also been reported \citep{riebe:2004, maher:2014, krissansen-totton:2017}. By default, we adopt $\beta_L=\beta_S=0.3$. However, the dependence of weathering on the \co\ level is not well constrained, so we also consider other $\beta$ values below.

Most seafloor weathering occurs in young oceanic crust because older crust tends to be covered in a thick layer of sediment that limits the degree of interaction with seawater \citep{coogan:2015, coogan:2018}. Thus, the seafloor weathering rate is limited by the supply of fresh oceanic crust, and is proportional to the seafloor spreading rate. The spreading rate is assumed to be proportional to the outgassing rate $G$ \citep{sleep:2001}. Conversely, land-based weathering is independent of the outgassing rate. 

Seafloor weathering depends on ocean acidity, which depends on atmospheric \co\ \citep{abbot:2012, krissansen-totton:2017, krissansen-totton:2018, schwieterman:2019}. We model this by simply assuming that seafloor weathering has a power law dependence on the atmospheric \co\ partial pressure $P$. Newly created seafloor is only weatherable for a finite amount of time due to sedimentation, so $W_S$ in Eqn.~\ref{eq-weathering} depends on both the spreading rate and the rate at which the new crust is weathered.

The seafloor weathering rate depends on the temperature of the rock beneath the ocean rather than the surface temperature. \citet{krissansen-totton:2017} estimate that the temperature at the base on the ocean is currently about 11 K lower than the surface temperature, while the mean temperature of the weatherable rock beneath the seafloor is about 9 K higher than the temperature at the base of the ocean. Thus, these two effects largely cancel each other, and we will assume that the seafloor weathering temperature is the same as the surface temperature, or 10 K above freezing, whichever is higher:
\begin{equation}
\tsf=\max(T,{\rm 283\ K})
\end{equation}

During globally glaciated epochs, land based weathering is thought to be negligible. However, seafloor weathering will continue because the ocean is unlikely to freeze at all depths due to heat flowing from the mantle \citep{nakayama:2019},. Continued mixing of \co\ between the atmosphere and oceans is also likely \citep{lehir:2008}, so seafloor weathering should continue to influence the climate. We will revisit these assumptions in Section~4. In globally glaciated cases, we calculate the seafloor weathering rate assuming that the ocean base is at 273 K, and so the mean seafloor weathering temperature is $\tsf=283$ K.

There is considerable uncertainty regarding how the outgassing rate varies over time on Earth-like planets, although it is plausible that outgassing declines over time as the mantle cools. Here we consider two possibilities: (i) a fixed outgassing rate, and (ii) an exponentially declining outgassing rate that matches the current rate on Earth after 4.5~Gy. In the latter case, the outgassing rate $G$ at time $t$ is
\begin{equation}
G=G_0\exp(-t/\tau)
\end{equation}
where $G_0$ is the initial outgassing rate, and
\begin{equation}
\tau=\frac{\rm 4.5\ Gy}{\ln G_0}
\end{equation}

Using the climate and weathering models described above, we search for equilibrium values of $T$ and $P$ assuming a single globally averaged surface temperature. As we will see in the following sections, there are some cases with more than one equilibrium state, and there are also situations where no equilibrium exists. We discuss each of these cases as they arise.

%
%
\section{Results}
\subsection{Land Weathering Only}
We begin by considering an Earth-like planet subject to land-based weathering only. We consider two cases: (i) a fixed outgassing rate $G$ similar to that on Earth today, and (ii) a fixed outgassing rate 5 times higher than this. We use the land-based weathering parameters $\theta_L=10$ K, $\beta_L=0.3$, and an intrinsic weathering rate $W_{L0}=1$. 

Figure 1 shows the equilibrium temperature and \co\ partial pressure as a function of solar insolation $S$ normalized to that on Earth today. The red curves in the figure show the case with $G=1$ case, and the blue curves show the case with $G=5$.

\begin{figure}
\plotone{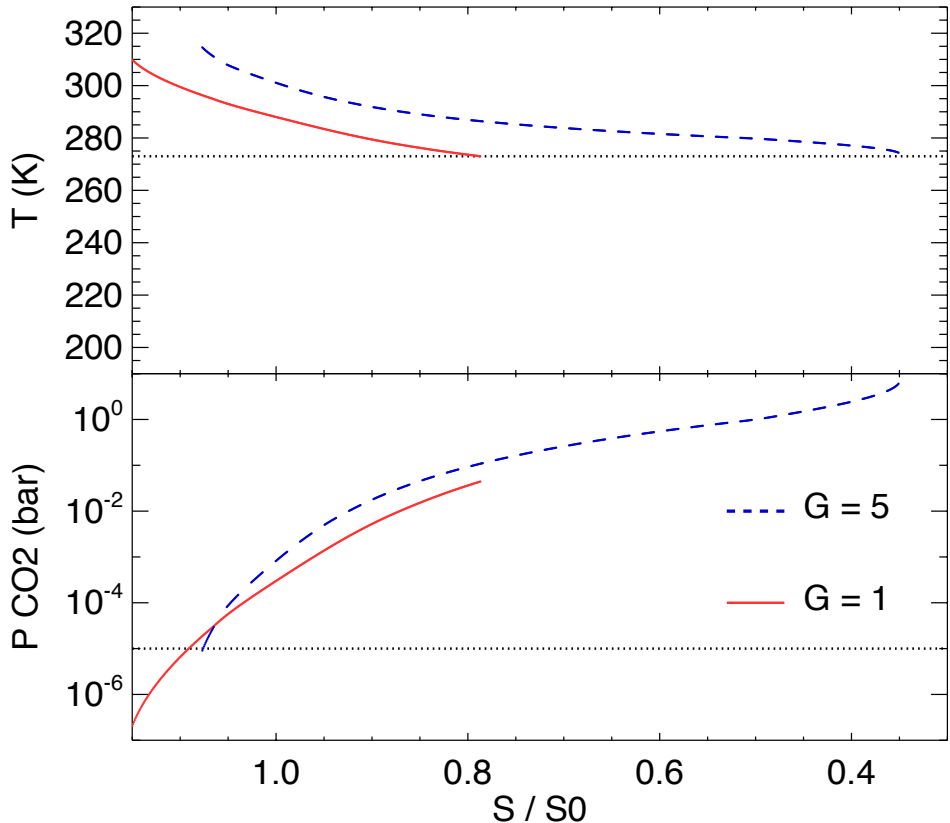}
\caption{Equilibrium global temperature and \co\ partial pressure for an Earth-like planet as a function of stellar insolation $S$ for land-based weathering only. The outgassing rate is fixed at the current rate for Earth (red curves), and 5 times the current rate (blue curves). Note that no equilibrium exists for the red curves when $S/S_0<0.78$.}
\end{figure}

For the higher outgassing rate, the coupled combination of weathering and climate establishes a stable climate with a temperature that depends modestly on insolation. The temperature varies from just above freezing at $S/S_0=0.35$ to about 320 K at $S/S_0=1.08$. The \co\ partial pressure varies by about 7 orders of magnitude over this same range of $S$ values, with consequent changes in the greenhouse warming. This shows how the climate-weathering feedback operates to moderate the change in temperature that would be driven by varying insolation otherwise.

For insolations $S/S_0>0.78$, the case with $G=1$ behaves similarly, albeit with lower $T$ and $P$ for a given insolation. Since the outgassing rate is lower, the equilibrium weathering rate is also lower, so $T$ and $P$ do not need to be as large as the previous case in order for weathering to balance the outgassing rate. 

\begin{figure}
\plotone{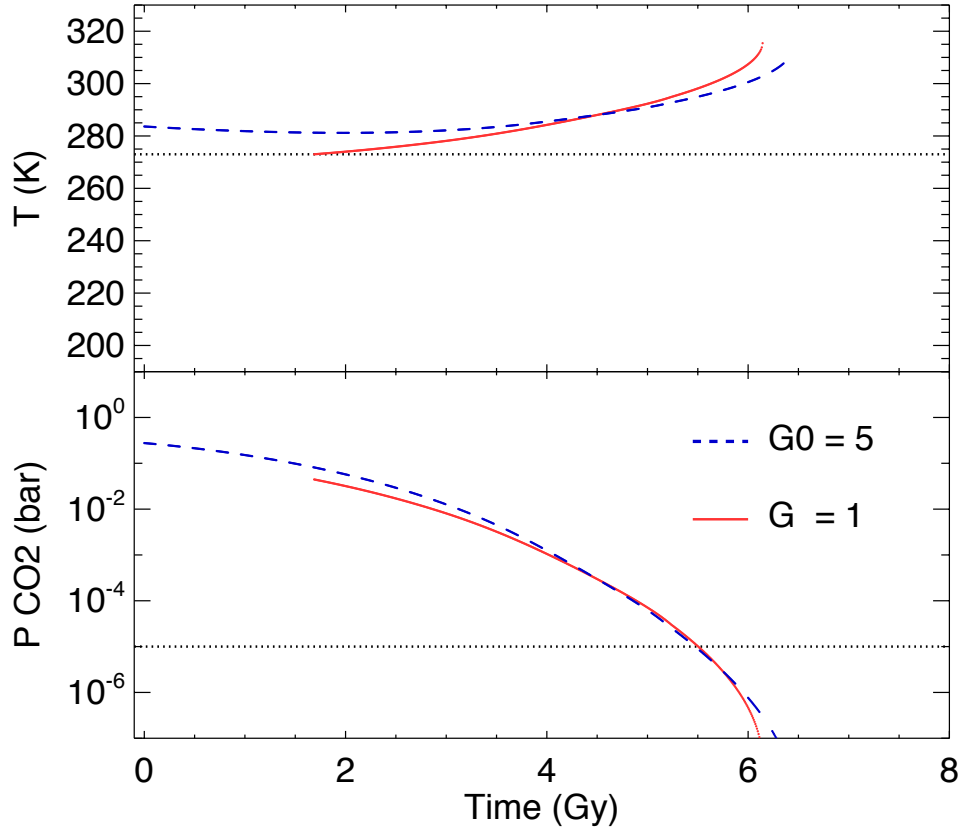}
\caption{Equilibrium global temperature and \co\ partial pressure for an Earth-like planet as a function of time for land-based weathering only with $W_{L0}=1$. The outgassing rate is fixed at the current rate for Earth (red curves), or decays exponentially with an initial value $G_0=5$ times the current rate (blue curves).}
\end{figure}

For $S/S_0<0.78$ the behavior is different. At temperatures just above freezing, the \co\ level is high enough that the weathering rate exceeds the outgassing rate. At slightly lower $T$, the surface freezes and land-based weathering ceases. This discontinuity means that no equilibrium configuration exists. Time varying models show that in cases like this the climate oscillates between a globally frozen state with no weathering, and a hot, ice-free state with rapid weathering \citep{menou:2015}.  This kind of ``limit cycle'' solution typically occurs in situations with low outgassing rates, or when the intrinsic weathering rate is high \citep{mills:2011, abbot:2016}.

In principle, we could imagine an equilibrium in which the planet's surface is only partially covered in ice, thereby reducing the fraction of land that undergoes weathering until it matches the outgassing rate. However, models that consider latitudinal variation of temperature find that partially ice-covered solutions tend to be unstable for low-obliquity planets like Earth \citep{caldeira:1992a, spiegel:2008, haqq-misra:2016}. Small perturbations lead to rapid, runaway advance or retreat of the ice-line latitude due to a strong feedback between temperature, ice coverage and albedo. The end result is a planet entirely ice covered or ice free (possibly retaining small ice caps at the poles). Here, we will assume that such an instability occurs, so that the climate does not remain in a partially ice-covered equilibrium.

The results in Figure 1 are similar to Figure~2 of \citet{kadoya:2019}. These authors considered the same two outgassing rates as here but in a model that divides the planet into a series of latitudinal bands with temperature and weathering calculated at each latitude. This similarity suggests that the simpler global model used here can offer a rough guide to climate behavior. 

From the rest of this section, we will consider the evolution of a planet at a fixed distance from the star as the star's luminosity increases over time. Figure 2 shows the evolution of a planet at 1 AU subject solely to land-based weathering for the same weathering parameters used in Figure 1. The red curves show a case with fixed outgassing at a rate equal to that on Earth today. The blue curves show an exponentially decaying outgassing rate, with initial value $G_0=5$.

For the fixed-outgassing case (red curves), the balance between weathering and outgassing provides long-term climate stability as expected. The global temperature increases over time and \co\ pressure decreases in order to maintain equilibrium as the star grows more luminous. However, no equilibrium exists for the first 1.6 Gy of the planet's history. Instead, the planet would undergo glaciation limit cycles prior to 1.6 Gy.

The evolution is somewhat different for the case with exponentially declining outgassing rate (blue curves). Early on, the climate is above freezing, and $T$ actually declines slightly over time. The insolation increases with time but this effect is more than offset by the rapid early decline in the outgassing rate. After about 2 Gy, the change in insolation becomes more important, and $T$ begins to increase again. Unlike the fixed-outgassing case, the variable-outgassing case maintains an equilibrium climate above freezing for all times until the equilibrium state disappears at about 6.4 Gy. At this point, the climate is likely to enter a runaway greenhouse with no stable equilibrium until the surface becomes hot enough to radiate at visible wavelengths \citep{kasting:1993}.

\begin{figure}
\plotone{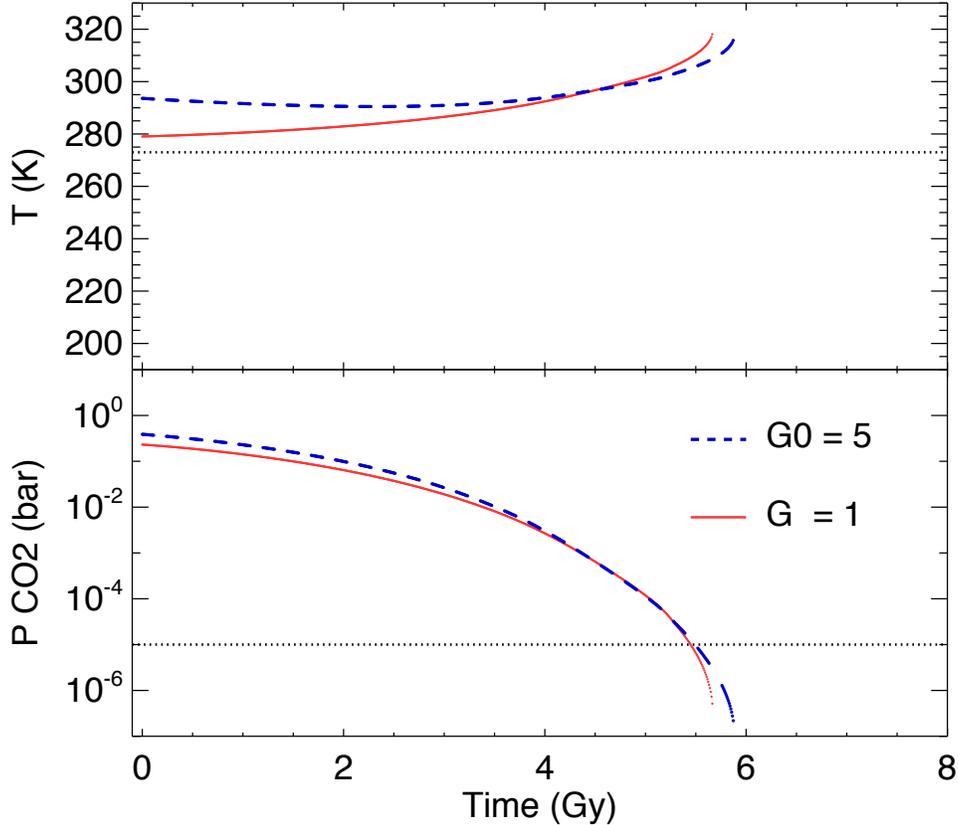}
\caption{Equilibrium global temperature and \co\ partial pressure for an Earth-like planet as a function of time for land-based weathering only with $W_{L0}=1/3$. The outgassing rate is fixed at the current rate for Earth (red curves), or decays exponentially with an initial value $G_0=5$ times the current rate (blue curves).}
\end{figure}

Overall, the temperature in the variable-outgassing model stays within a relatively narrow band throughout the evolution until the runaway greenhouse state begins. The declining outgassing rate over time acts to offset the rising insolation to some degree, leading to less climate variability than the case with constant outgassing.

The planet becomes uninhabitable for advanced life, based on our definition, at about 5.4 Gy in both cases, when the \co\ partial pressure drops below $10^{-5}$ bar. The temperature is still relatively clement at this point, well below the 323 K threshhold we have adopted here. In this case, low \co\ levels are likely to constrain the existence of advanced life before high temperatures do. This is the typical outcome for the cases we consider here.

On the modern Earth, continental weathering rates are probably enhanced by a factor of a few due to the presence of land-based planets \citep{schwartzman:1989, kump:2000, lenton:2001}. Before plants existed on land, weathering rates would have been lower as a result. Weathering rates would also be lower on planets without land plants, other things being equal. We consider this effect now.

Figure 3 shows a planet at 1 AU with land weathering only, and weathering rates reduced by a factor of 3 compared to case shown in Figure~2, so that $W_{L0}=1/3$. Other model parameters are the same in both figures.

A few differences from the previous case are notable. Equilibrium temperatures are higher in Figure~3 than Figure~2, typically by about 10 K. This difference arises because the intrinsic weathering rate is lower in Figure~3, so $T$ and/or $P$ must be larger to match the outgassing rate. In practice, $T$ and $P$ are positively correlated due to the greenhouse effect so both increase, although the difference is more obvious for $T$ in the figures. The higher equilibrium temperatures in Figure~3 mean that stable ice-free climates exist for both outgassing models for all times until roughly 5.7 Gy (fixed outgassing case) or 5.9 Gy (variable outgassing case).

\begin{figure}
\plotone{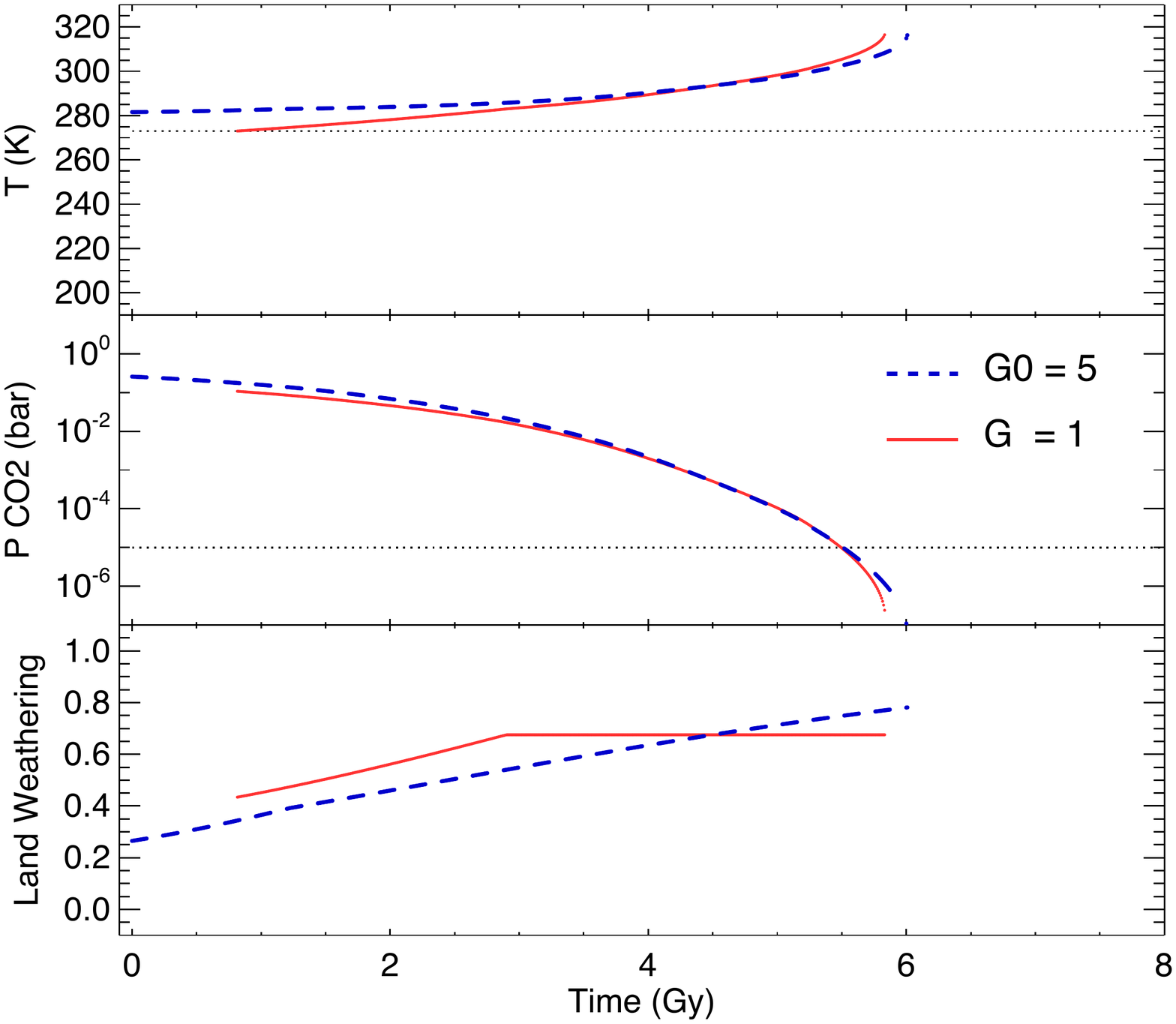}
\caption{Equilibrium global temperature and \co\ partial pressure for an Earth-like planet as a function of time with land-based and seafloor weathering, with $W_{L0}=1/3$ and $W_{S0}=1/6$. The outgassing rate is fixed at the current rate for Earth (red curves), or decays exponentially with an initial value $G_0=5$ times the current rate (blue curves).}
\end{figure}

%
%
\subsection{Land and Seafloor Weathering}
We now introduce seafloor weathering in addition to land-based weathering. The rates of continental and seafloor weathering are likely to vary from one planet to another. The relative importance of the two kinds of weathering may also vary. Differences in the surface land fraction, the chemical composition of continental and seafloor rocks, the presence and distribution of biological activity, precipitation and sedimentation rates, and ocean chemistry can all affect weathering rates. Given these uncertainties, we will examine several different values for the intrinsic weathering rates $W_{L0}$ and $W_{S0}$, beginning with values that lead to qualitatively different outcomes. We will survey different weathering rates more broadly in Section~3.5.

Figure 4 shows the climate evolution in a case similar to Figure~3 except that both land and seafloor weathering operate with $W_{L0}=1/3$ and $W_{S0}=1/6$. Thus, at the current temperature, \co\ partial pressure and outgassing rate on Earth, the total weathering rate would be half that observed today. The red curves show a planet with constant outgassing with $G=1$, while the blue curves show an exponentially declining outgassing rate with an initial value of $G_0=5$. In addition to the temperature $T$, and \co\ partial pressure $P$, Figure~4 also shows the fraction of the total weathering rate that occurs on land as a function of time.

In the fixed-outgassing case (red curves), the increased intrinsic weathering rate compared to Figure~3 means that no equilibrium state exists at early times in Figure~4. When the surface is covered in ice, land weathering ceases. Seafloor weathering continues, but it is insufficient to balance the outgassing rate, so \co\ builds up in the atmosphere until temperatures rise above freezing. At this point, land based weathering begins, and the total weathering rate immediately becomes high enough to start reducing \co\ levels. As a result, the planet refreezes, and the climate will cycle between frozen and ice-free states.

After about 0.7 Gy, the insolation has increased enough for an ice-free equilibrium state to appear. From this point on, the evolution is qualitatively similar to Figure~3, with somewhat lower $T$ due to the increase weathering rate.

In the variable-outgassing case (blue curves), an ice-free equilibrium state exists from the beginning. The temperature is lower than the same case in Figure~3, again due to the increased weathering rate. The decrease in $T$ at early times in Figure~3 is not seen in Figure~4. This is because of the contribution of seafloor weathering, which occurs at a rate proportional to the outgassing rate in our model. The presence of seafloor weathering moderates the effect of high outgassing rates at early times because the seafloor weathering rate is also higher by the same amount.

The last panel of Figure 4 shows the fraction of weathering due to land weathering. The adopted temperature and \co-pressure dependencies for the two weathering rates are the same in this case, but the relative importance of land weathering can vary for two reasons. Firstly, land weathering is independent of the outgassing rate, while seafloor weathering and outgassing are linked. This explains the gradual increase in the importance of land weathering when outgassing declines over time, shown by the blue curve. 

When the outgassing rate is fixed (red curves), land based weathering can still become increasingly important over time if the surface temperature $T$ is below 283 K. For $T<283$ K, the seafloor weathering temperature $\tsf$ is held at 283 K by heat escaping from the planet's interior. Thus, seafloor weathering becomes progressively less important than land weathering for increasing surface temperature. When $T>283$ K, the ratio of the two weathering rates becomes fixed, and the red curve in the last panel of Figure 4 becomes flat.

Figure 5 shows the effect of doubling the seafloor weathering compared to Figure~4, so that $W_{L0}=W_{S0}=1/3$. Thus for nominal values of $T$, $P$ and $G$, seafloor weathering is equally as effective as land weathering. This change leads to several qualitative differences from the previous examples.

\begin{figure}
\plotone{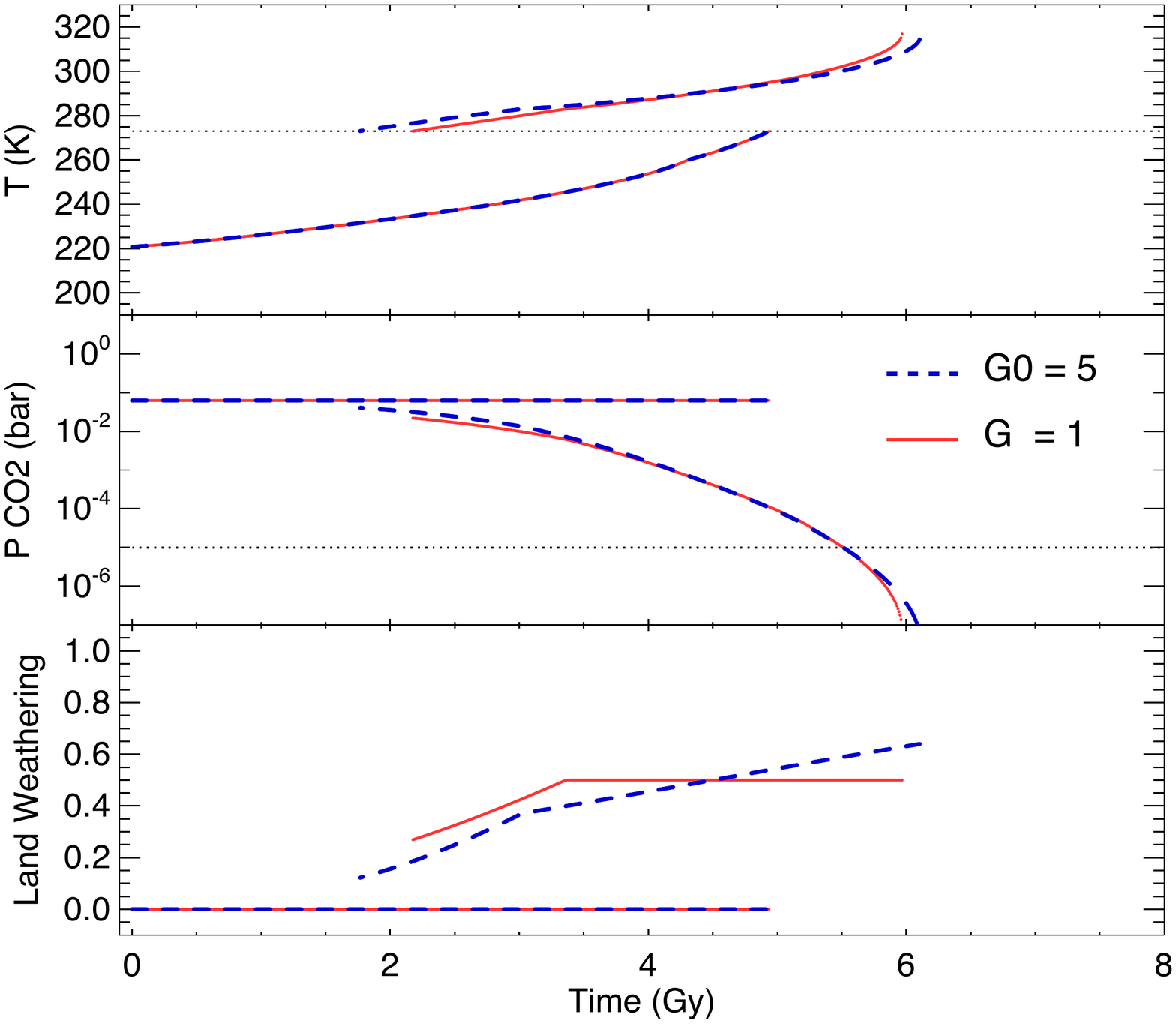}
\caption{Equilibrium global temperature and \co\ partial pressure for an Earth-like planet as a function of time with land-based and seafloor weathering, with $W_{L0}=W_{S0}=1/3$. The outgassing rate is fixed at the current rate for Earth (red curves), or decays exponentially with an initial value $G_0=5$ times the current rate (blue curves). Note that the two cases are identical for the first 1.7 Gy.}
\end{figure}

For the first 1.7 Gy of the planet's history, an equilibrium state exists that is marked by below-freezing temperatures and a surface covered in ice. There is no land-based weathering at this stage, but seafloor weathering operates and is effective enough to balance outgassing. The evolution is identical for the two outgassing models. The climate variables are the same for the two outgassing models. Seafloor weathering, which is the only weathering that operates, scales with the outgassing rate $G$, so the resulting equilibrium is independent of $G$.

The partial pressure of \co\ remains constant over time while this glaciated equilibrium exists. The seafloor weathering temperature is fixed at 283 K regardless of the surface temperature, so the weathering rate depends only on $P$ according to Eqn.~\ref{eq-weathering}. Since the weathering-to-outgassing ratio is constant for both outgassing models, $P$ also remains constant. As a result, $T$ rises quite rapidly over time, because no temperature-dependent negative feedback operates to stabilize the climate in this state.

At about 1.8 Gy (variable outgassing case), or 2.1 Gy (fixed outgassing case), a second, ice-free equilibrium state appears. The insolation has increased to the point where an ice-free climate has a low enough $P$ that outgassing can balance weathering.

For the next 2.5--3 Gy, the two equilibrium states coexist. The \co\ level is lower in the ice-free state because the albedo is lower and the insolation is balanced by a weaker greenhouse effect. As a result, two different equilibrium values of $P$ exist. In the high-$P$ case, outgassing is balanced solely by seafloor weathering. For the low-$P$ case, outgassing is offset by both land and seafloor weathering. 

Because the planet is in the glaciated equilibrium state prior to the appearance of the ice-free equilibrium, it seems likely that it will continue to follow the ice-covered trajectory when the second equilibrium appears. Thus, the planet will be uninhabitable according to the usual definition even though a habitable solution exists. However, given a large enough perturbation the climate could switch to the ice-free state and follow the other trajectory. 

After about 4.9 Gy, the ice-covered equilibrium disappears. If the climate was in the glaciated state up until this point it will transition abruptly to the ice-free equilibrium. This involves a substantial jump in $T$ and $P$ as a result, for either outgassing model. After this transition, the climate evolves in a similar way to the previously considered cases.

In Figure 6, we consider a case with even stronger seafloor weathering. Here, $W_{L0}=1/3$ as before, but now $W_{S0}=2/3$. For the first 3 Gy of evolution, the climate behaves in a similar way to the early evolution in Figure~5. A globally-ice-covered equilibrium state exists due to a balance between seafloor weathering and outgassing. This state lasts for longer than in Figure~5, and the corresponding surface temperature is lower, because weathering is more effective for a given partial pressure of \co. This means that $P$ is lower at equilibrium, and greenhouse warming is weaker.

\begin{figure}
\plotone{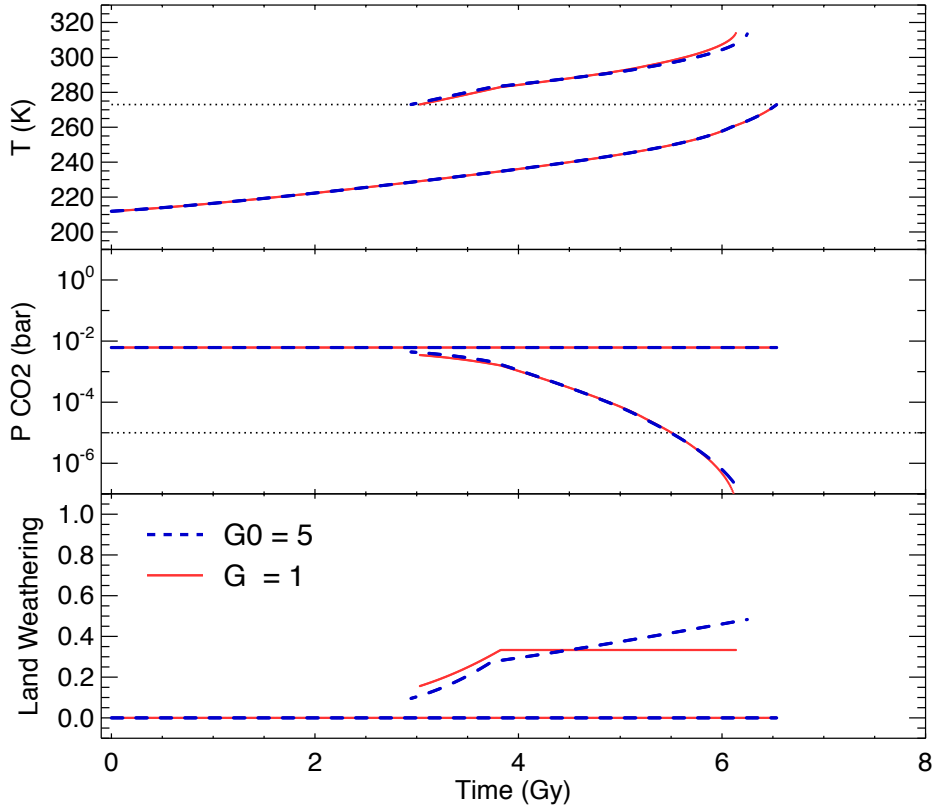}
\caption{Equilibrium global temperature and \co\ partial pressure for an Earth-like planet as a function of time with land-based and seafloor weathering, with $W_{L0}=1/3$ and $W_{S0}=2/3$. The outgassing rate is fixed at the current rate for Earth (red curves), or decays exponentially with an initial value $G_0=5$ times the current rate (blue curves). Note that the two cases are identical for the first 2.9 Gy.}
\end{figure}

At about 3 Gy, an ice-free equilibrium state appears, at slightly different times in the two outgassing models. From this point onwards, the ice-covered and ice-free equilibria evolve in a qualitatively similar way to the corresponding cases in Figure~5. At any given time, the equilibrium values of $P$ and $T$ are somewhat lower in Figure~6 due to the stronger intrinsic  weathering. As before, the evolution in the glaciated state is independent of the outgassing rate. The evolutions in the ice-free state for the two outgassing models are also similar to each other since this state is centered around 4.5 Gy, at which point $G$ is the same in both models by design.

Unlike Figure~5, in the case shown in Figure~6 the ice-free equilibrium ceases to exist slightly earlier than the glaciated equilibrium state. If the planet's climate avoids large perturbations it will remain in an ice-covered state for more than 6 Gy, after which the climate will transition directly to a runaway greenhouse state. Even though a habitable equilibrium climate is possible for roughly 3 Gy of the planet's history, the planet may never be habitable.

%
%
\subsection{Weathering $P$ and $T$ Dependence}
The parameters that determine the weathering rate are uncertain, even for Earth. This is particularly true for the \co-pressure dependences of land and seafloor weathering, denoted by $\beta_L$ and $\beta_S$ in Eqn.~\ref{eq-weathering}. In this section, we examine the effect of varying these parameters as well as the corresponding temperature dependencies $\theta_L$ and $\theta_S$.

Figure 7 shows the effect of different weathering \co-pressure dependencies on the climate evolution of an Earth-like planet at 1 AU. The green curves show a case with $\beta_S=\beta_L=0.1$, while the purple curves show a case where $\beta_S=\beta_L=0.5$. In both cases we use $W_{S0}=W_{L0}=1/3$, and we use an exponentially declining outgassing rate with $G_0=5$. The results can be compared with the blue curves in Figure~5 that show a case with identical  model parameters except that $\beta_S=\beta_L=0.3$.

\begin{figure}
\plotone{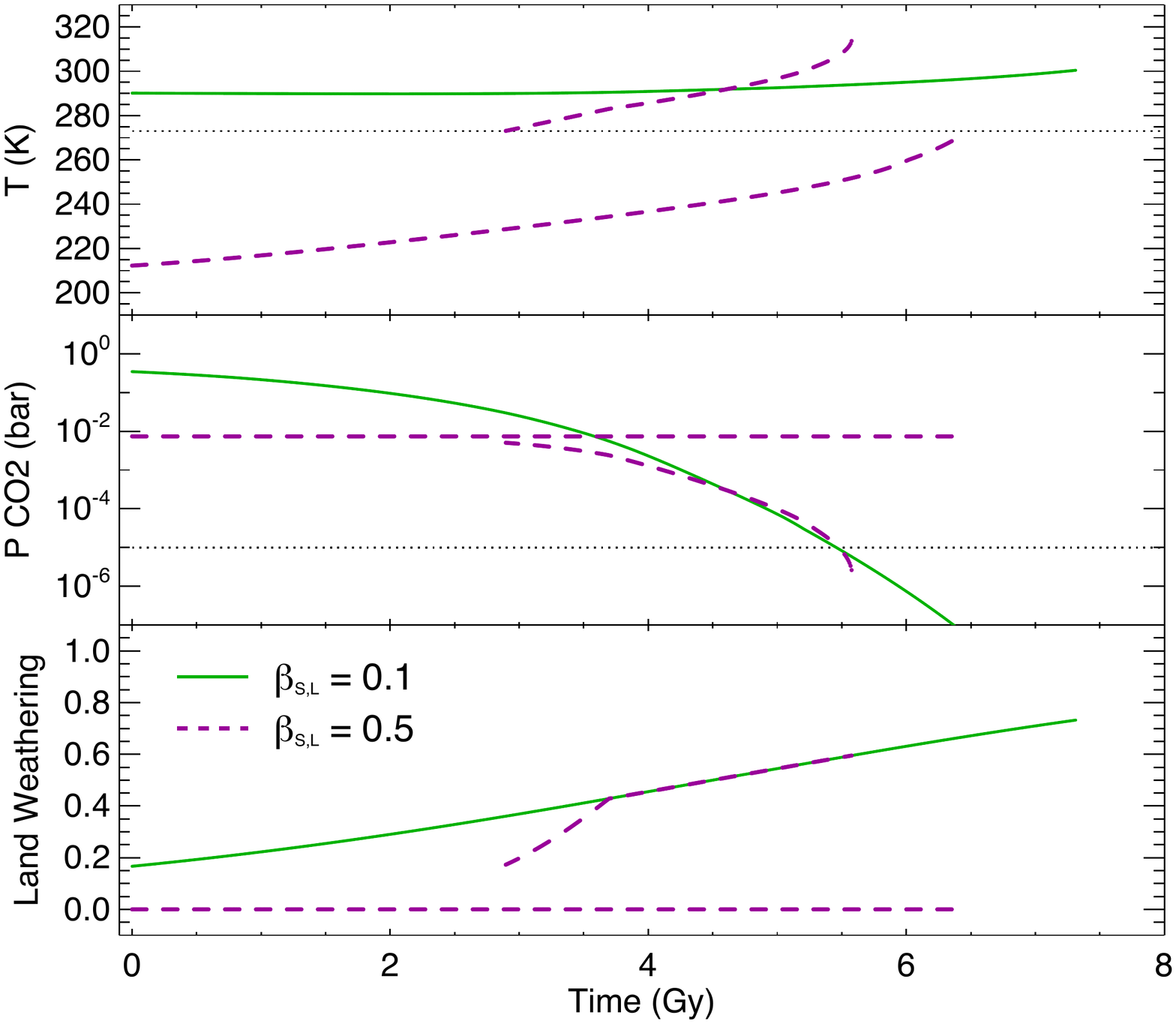}
\caption{Equilibrium global temperature and \co\ partial pressure for an Earth-like planet as a function of time with land-based and seafloor weathering, with $W_{L0}=W_{S0}=1/3$. The outgassing rate decays exponentially with an initial value $G_0=5$ times the current rate.  The green curves show a case with weathering \co-pressure dependences $\beta_L=\beta_S=0.1$. The purple curves show a case with $\beta_L=\beta_S=0.5$.}
\end{figure}

The \co-pressure dependence of the weathering rate clearly plays an important role in determining the equilibrium climate. When $\beta=0.1$, the temperature remains almost constant, remaining between roughly 290 and 300 K for more than 7 Gy. The very weak dependence of weathering on $P$ implies that the weathering rate depends almost entirely on temperature. This  means that the temperature cannot vary much for a given outgassing rate if weathering and outgassing are to balance. The variation in $T$ would be a little larger if $G$ were constant, due to changing insolation. However,  the fact that insolation and outgassing vary in opposite directions further acts to limit the variation in temperature.

The case with $\beta=0.5$ is very different. Now the weathering rate depends more strongly on \co\ pressure than the case shown in Figure~5, although the relation is still less than linear. The larger $\beta$ values result in a stronger variation in temperature with time when the planet is in an ice-free state. For the first 2.9 Gy of evolution in Figure~7, the only equilibrium state is one with $T$ below freezing. A second, ice-free equilibrium appears at 2.9 Gy, and the two states coexist until 5.6 Gy. At this point, the ice-free solution disappears while the glaciated case persists until about 6.4 Gy. 

Overall, the purple curves in Figure~7 resemble the evolution in Figure 6 which considered a case with stronger intrinsic seafloor weathering but a weaker \co-pressure dependence. In the glaciated equilibrium, $P$ remains constant while $T$ increases quite rapidly over time. In both figures, the glaciated state remains after the ice-free equilibrium disappears, however the difference is clearer in Figure 7 where the ice-covered state outlives the ice-free one by more than 0.8 Gy.

Looking at the two cases in Figure~7 and the blue curves in Figure~5 we see a clear trend in the time at which the habitable, ice-free state vanishes due to rising insolation. When $\beta=0.5$, the ice-free equilibrium disappears at 5.6 Gy. For $\beta=0.3$, this state lasts until 6.1 Gy, while for $\beta=0.1$, the planet maintains an ice-free equilibrium state until about 7.3 Gy. Clearly, a weathering rate that depends only weakly on the \co\ pressure is an important factor for maintaining a long-lived biosphere.

\begin{figure}
\plotone{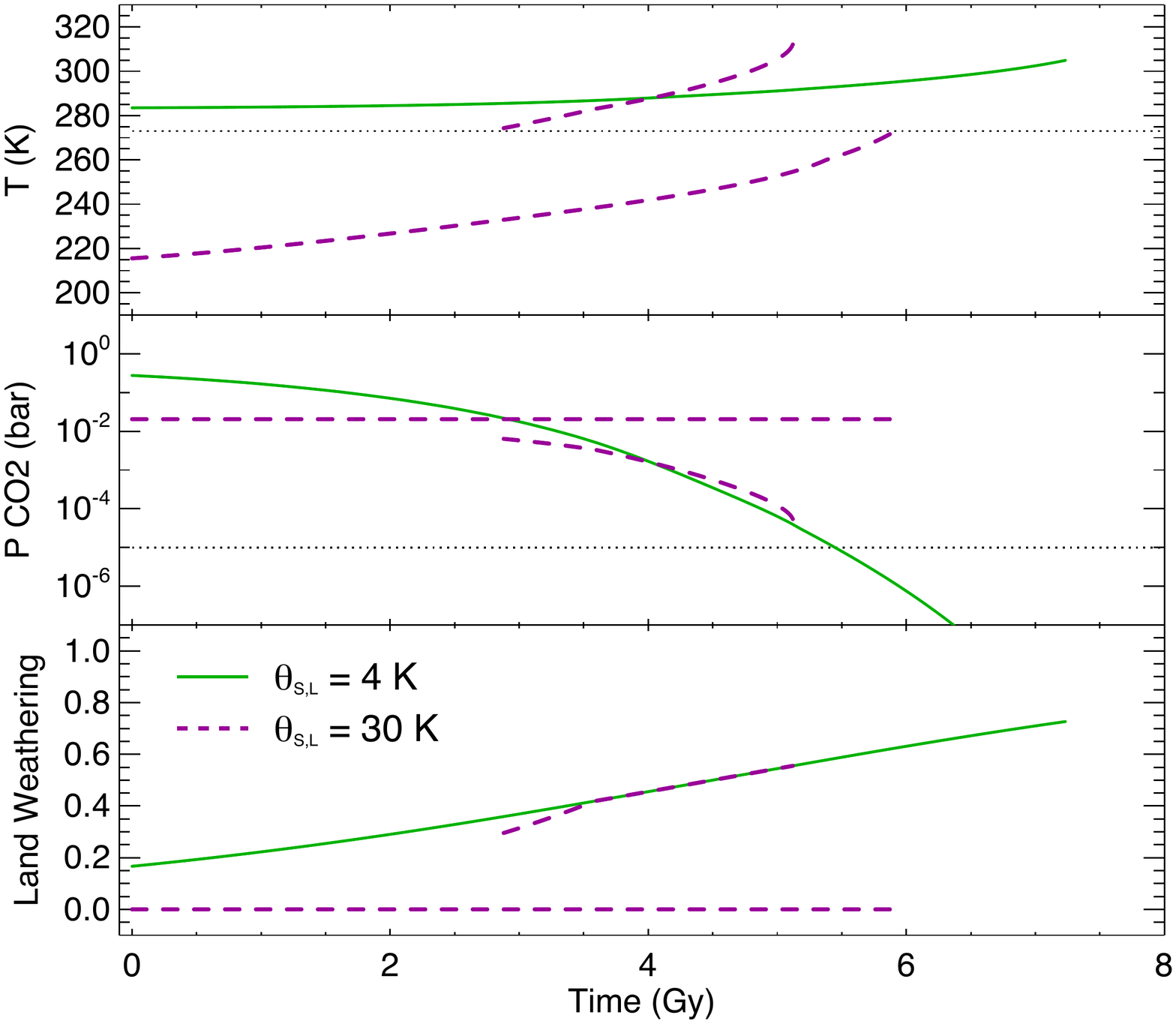}
\caption{Equilibrium global temperature and \co\ partial pressure for an Earth-like planet as a function of time with land-based and seafloor weathering, with $W_{L0}=W_{S0}=1/3$. The outgassing rate decays exponentially with an initial value $G_0=5$ times the current rate. The green curves show a case with weathering temperature dependence $\theta_L=\theta_S=4$ K. The purple curves show a case with $\theta_L=\theta_S=30$ K.}
\end{figure}

Figure 8 shows the effect of varying the land and seafloor weathering temperature dependence. The green curves show a case with $\theta_L=\theta_S=4$ K, while for the purple curves we use $\theta_L=\theta_S=30$ K. Both cases use $\beta_L=\beta_S=0.3$.  Note that we have deliberately included a small value for the temperature dependence parameters, which probably falls outside the plausible range of values, in order to clearly show the effect of varying these parameters.

Overall, the behavior seen in Figure 8 resembles that in Figure~7. Small values of the temperature parameters, $\theta_L$ and $\theta_S$, result in a stable, ice-free climate with relatively small variation in temperature. Large values of $\theta$ allow a glaciated state to exist, and lead to a short-lived ice-free equilibrium that exhibits a rapid rise in $T$ over time.

Changing the \co-pressure or the temperature weathering sensitivity has a similar effect on the climate in the ice-free state. We can see the reason for this by considering a highly simplified model for climates similar to that on Earth today. In this model, the outgoing infrared radiation depends logarithmically on the \co\ partial pressure due to saturation of the \co\ absorption bands \citep{abbot:2012}. For small variations in temperature we can also linearize the relation between outgoing radiation and $T$. Thus we have
\begin{equation}
\frac{(S-S_0)}{4}(1-\alpha)=K(T-T_0)-J\phi
\end{equation}
where $J\simeq 10$ W/m$^2$ and $K\simeq 2$ W/m$^2$/K \citep{abbot:2016}, $S$ is the insolation, $\alpha$ is the albedo (which we assume is constant) and
\begin{equation}
\phi=\ln\frac{P}{P_0}
\end{equation}

We will assume, for simplicity that land and seafloor weathering have the same \co-pressure and temperature dependence. At equilibrium we have
\begin{equation}
G=(W_{L0}+GW_{S0})\exp\left[\frac{T-T_0}{\theta}+\beta\phi\right]
\end{equation}
where $G$ is the outgassing rate. Eliminating $\phi$, we get
\begin{equation}
(T-T_0)
=\left[\frac{J\theta}{J+K\beta\theta}\right]
\ln\left(\frac{G}{W_{L0}+GW_{S0}}\right)
+\left[\frac{\beta\theta S_0(1-\alpha)}
{4(J+K\beta\theta)}\right]
\frac{(S-S_0)}{S_0}
\end{equation}

This expression gives the equilibrium temperature $T-T_0$ as a function of the normalized insolation $(S-S_0)/S_0$, and thus time. ($G$ also varies with time. However, the temperature depends on $G$ only logarithmically, and the first factor in square brackets is smaller than the second.) We note that the coefficient of the insolation term  depends on the product of $\beta$ and $\theta$. To a first approximation, the rate of change of the equilibrium temperature (and \co\ pressure) will depend on $\beta\times\theta$ rather than these quantities individually. Thus, changing $\theta$ has roughly the same effect on the climate as changing $\beta$ for an ice-free climate similar to Earth.

%
%
\subsection{Dependence on Planetary Distance}
Until this point we have considered planets orbiting 1 AU from a Sun-like star. In this  section we look at planets at other orbital distances. Rather than examining individual climate trajectories, we will map out the different climate regimes---ice-free equilibrium, glaciated equilibrium, limit cycles etc.---as a function of distance and time.

\begin{figure}
\plotone{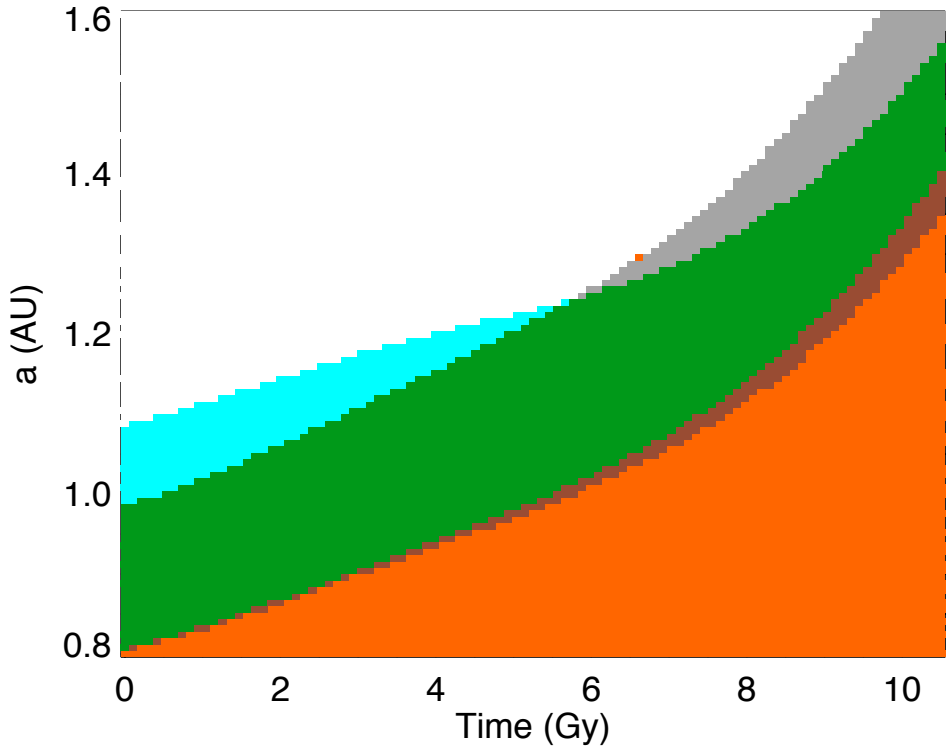}
\caption{The different climate regimes as a function of time and planetary distance $a$. Planets in the white region are globally glaciated, those in the green area are ice-free and habitable, and those in the orange region are undergoing a runaway greenhouse. Both glaciated and ice-free equilibrium states exist in the cyan region, and neither equilibrium exists in the grey area. The brown region contains planets that have habitable temperatures, but \co\ pressures $<10^{-5}$ bar. Model parameters are $W_{L0}=W_{S0}=1/6$, $\beta_L=\beta_S=0.3$, and $\theta_L=\theta_S=10$ K.}
\end{figure}

Figure 9 shows the climate states for weathering parameters $W_{S0}=W_{L0}=1/6$, and $\theta_S=\theta_L=10$ K, and $\beta_S=\beta_L=0.3$. Here we assume an exponentially decaying outgassing rate, initially 5 times that on Earth today.

The green and white regions show portions of phase space where a single equilibrium state exists, ice-free or glaciated respectively. The cyan region indicates that both of these equilibrium states exist, and the grey region indicates that neither equilibrium occurs so that the climate undergoes limit cycles. Planets in the orange region are undergoing a runaway greenhouse. Between the orange and green region is a narrow brown zone. Planets in this zone are in the ice-free equilibrium state, but the \co\ level is below $10^{-5}$ bar, which may be inhospitable for advanced life. 

In Figure 9 it is apparent that the age and orbital distance of a planet both play an important role in determining the climate regime. The various climate states typically occupy upward-sloping diagonal regions as a consequence of the increasing stellar luminosity over time. A planet at a fixed distance is likely to pass through 2 or more climate regimes during the main sequence lifetime of its star. 

For this set of model parameters, climates with a single equilibrium state make up substantial portions of phase space, while the regions with two stable states (cyan), or none (grey), make up smaller areas. The fraction of time that a planet exists in the ice free equilibrium with $P<10^{-5}$ bar (brown) is small in all cases. As we have seen, a planet that begins in the glaciated state (white) is likely to remain in this state if an ice-free equilibrium also appears, in the absence of a large perturbation. As a result, the cyan region may be considered an extension of the white region in practice.

\begin{figure}
\plotone{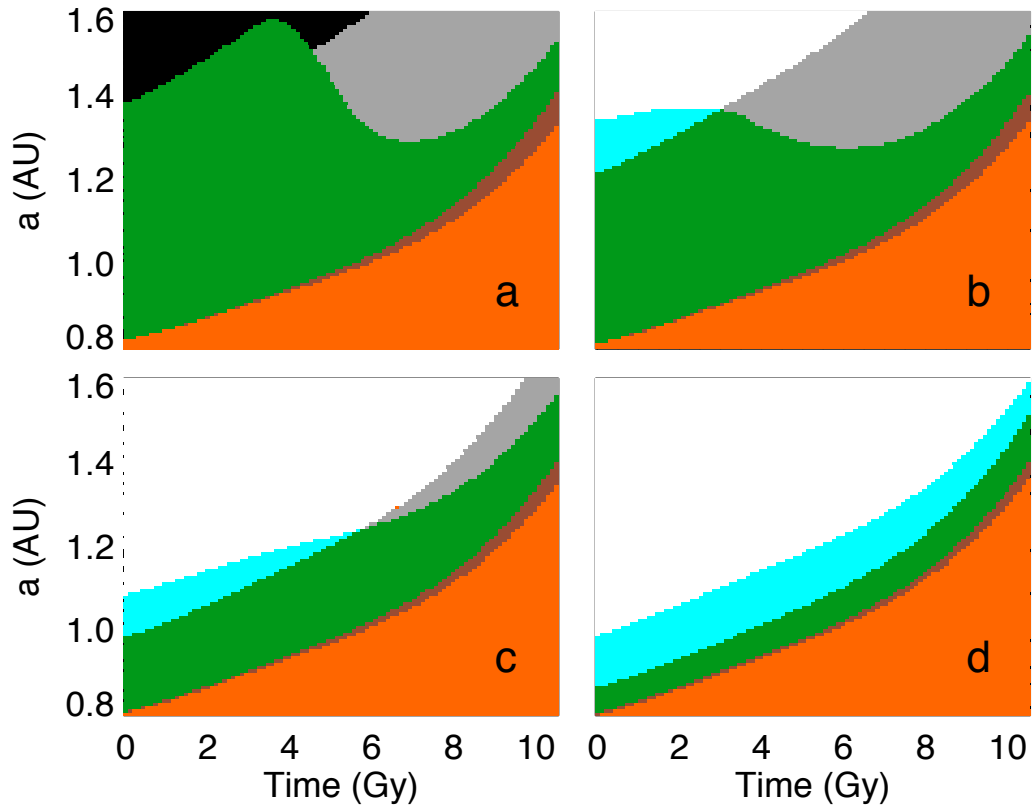}
\caption{Different climate states as a function of time and planetary distance $a$. The colors have the same meanings as in Figure~9. The four panels show differences in the intrinsic degree of land weathering and seafloor weathering, with the same total amount. Panels a, b, c, and d have $W_{L0}=1/3$, 1/4, 1/6 and 1/12 respectively. In all cases $W_{L0}+W_{S0}=1/3$. The values of the $\beta$ and $\theta$ parameters are the same as in Figure~9.}
\end{figure}

Although the ice-free equilibrium state (green) occupies a large region in Figure~9, it is by no means dominant. A planet at 1 AU will remain habitable for roughly 5.5 Gy, but the habitable lifespan can be substantially shorter for planets at other distances, especially for $r<0.9$ AU or $r>1.2$ AU. For this set of model parameters, the existence of the glaciated equilibrium (white) may significantly reduce the region that is conventionally considered habitable \citep{kasting:1993, forget:1997, kopparapu:2013}.

Figure 10 shows how the relative importance of land weathering and seafloor weathering affects the extent of the different climate states shown in Figure~9. In each of the four panels of Figure~10, the total intrinsic weathering rate is the same, such that $W_{L0}+W_{S0}=1/3$, but the ratio of these quantities differs. The other model parameters are the same as in Figure~9. Stable climates may occur in the black region of panel~a, but these will have \co\ levels above 10 bar, for which our climate model is not applicable.

Panel a of Figure 10 shows a case with land weathering only ($W_{L0}=1/3$ and $W_{S0}=0$). For these model parameters, no glaciated equilibrium exists at any time or distance. A stable, ice free-state covers a large fraction of phase space. For the first 4 Gy, the climate is habitable over a wide range of distances spanning about 0.6 AU. At later times, the ice-free equilibrium narrows considerably, and much of the phase space is occupied either by limit cycles (grey) or runaway greenhouse (orange). Towards the end of the star's main sequence evolution, an increasing fraction of the ice-free equilibrium state has very low \co\ partial pressures, indicated by the brown color. It is notable that a planet near 1.2 AU remains habitable for roughly 8 Gy.

The wave-like shape of the outer edge of the ice-free (green) state is caused by a combination of a declining outgassing rate and an increasing insolation over time. At equilibrium, weathering balances outgassing. Weathering depends on temperature and \co\ pressure, so $P$ and $T$ are linked. Prior to about 4 Gy, the outer edge of the ice-free equilibrium is set by the point at which $P=10$ bar, which is the maximum \co\ pressure that our climate model can simulate. The resulting equilibrium temperature is over 273 K, which suggests that an ice-free state could exist at somewhat larger distances with $P>10$ bar.

After about 4 Gy, the declining outgassing rate means that planets at large distances can no longer reach an equilibrium between outgassing and weathering. The outer edge of the ice-free (green) region is now the point at which the equilibrium temperature is just above freezing. For a given outgassing rate, there is a maximum value of $P$ such that weathering balances outgassing. Planets at a slightly larger distance will have a larger value of $P$, in order to offset the weaker insolation. For these planets, weathering will exceed outgassing at just above the freezing point, and limit cycles will occur.

The outer edge of the (green) ice-free region in Figure 10a moves inwards as the outgassing rate falls over time until about 6.5 Gy. At this point, the outgassing rate changes slowly with time, and changes in insolation become more important. The outer edge of the ice-free region now begins to move outwards as the star brightens.

Panels b, c and d of Figure 10 show progressively greater amounts of seafloor weathering with a corresponding decrease in land weathering. Several trends are apparent. A stable, glaciated equilibrium (white region) appears, and grows progressively larger as seafloor weathering becomes stronger. As a result, the ice-free equilibrium region shrinks. In panel d of the figure, 3/4 of intrinsic weathering is due to seafloor weathering, and the habitable (green) region has shrunk to a narrow strip that lasts for no more than 2 Gy.

The wavy outer edge of the ice-free state becomes less pronounced with each increase in seafloor weathering because seafloor weathering scales with the outgassing rate, negating the effect of the larger outgassing rate at early times. The behavior at the inner edge of the ice-free equilibrium is largely unaffected by the switch between land-based and seafloor weathering. The transition to the runaway state (orange) is almost the same in all four panels because at these high temperatures, the effects of land and seafloor weathering are similar apart from the dependence on $G$.

Although the intrinsic level of weathering is the same in all the panels of figure 10, the behavior is very different. This shows the importance of knowing whether weathering is dominated by land or seafloor reactions.  An increasing degree of seafloor weathering decreases the size of the habitable region for several reasons. Firstly it introduces the stable ice-covered state, which expands over more phase space as seafloor weathering becomes stronger. Secondly, introducing seafloor weathering reduces the importance of an early high rate of outgassing rate since weathering rates are also higher. Thirdly, for surface temperatures $273<T<283$ K, seafloor weathering is more effective than land based weathering for the same weathering parameters since the seafloor weathering temperature remains at 283 K.

\begin{figure}
\plotone{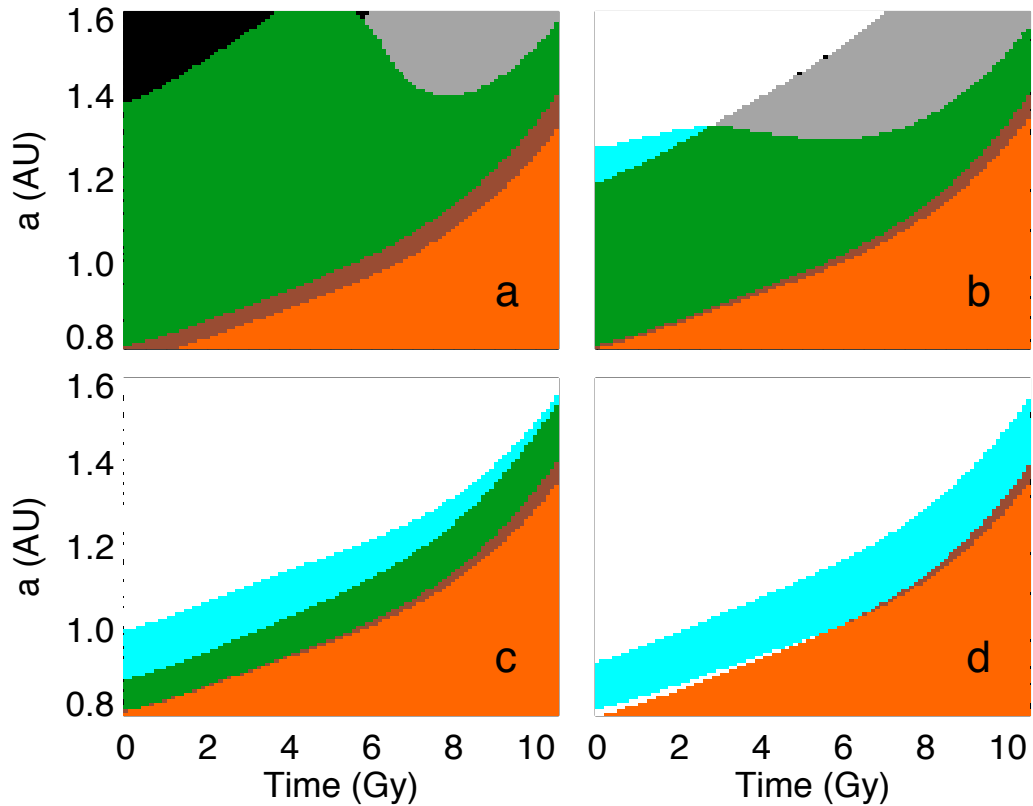}
\caption{Different climate states as a function of time and planetary distance $a$. The colors have the same meanings as in Figure~9.  The four panels show differences in the \co-pressure dependence of seafloor weathering. Panels a, b, c, and d have $\beta_S=0.1$, 0.25, 0.35 and 0.5 respectively. In all cases, $W_{L0}=W_{S0}=1/6$. Values of other model parameters are the same as in Figure~9.}
\end{figure}

We now look at the effect of varying the \co-pressure dependence of seafloor weathering on the extent of the different climate states. The 4 panels of Figure~11 show cases with $\beta_S=0.1$, 0.25, 0.35 and 0.5 respectively. The intrinsic land and seafloor weathering rates are equal, with $W_{L0}=W_{S0}=1/6$ in all 4 panels of the figure. The temperature and \co-pressure dependences of land-based weathering, and the temperature dependence of seafloor weathering are all kept fixed, with $\theta_L=\theta_S=10$ K, and $\beta_L=0.3$.

Overall, the behavior and the trends in Figure~11 are similar to those in Figure~10. A weak \co-pressure dependence for seafloor weathering results in a large ice-free climate region similar to the cases in Figure~10 with mainly land-based weathering. Increasing the seafloor \co-pressure dependence reduces the stabilizing effect of seafloor weathering on climate, reducing the ice-free region and leading to a larger portion of phase space covered by the glaciated state. This is the same behavior seen in Figure~10 as seafloor weathering is strengthened. In Figure~11, panel d, which has the strongest \co-pressure dependence, the ice-free equilibrium has shrunk to a tiny strip of phase space characterized by \co\ partial pressures below $10^{-5}$ bar. All regions of phase space are essentially uninhabitable for advanced life in this case.

\begin{figure}
\plotone{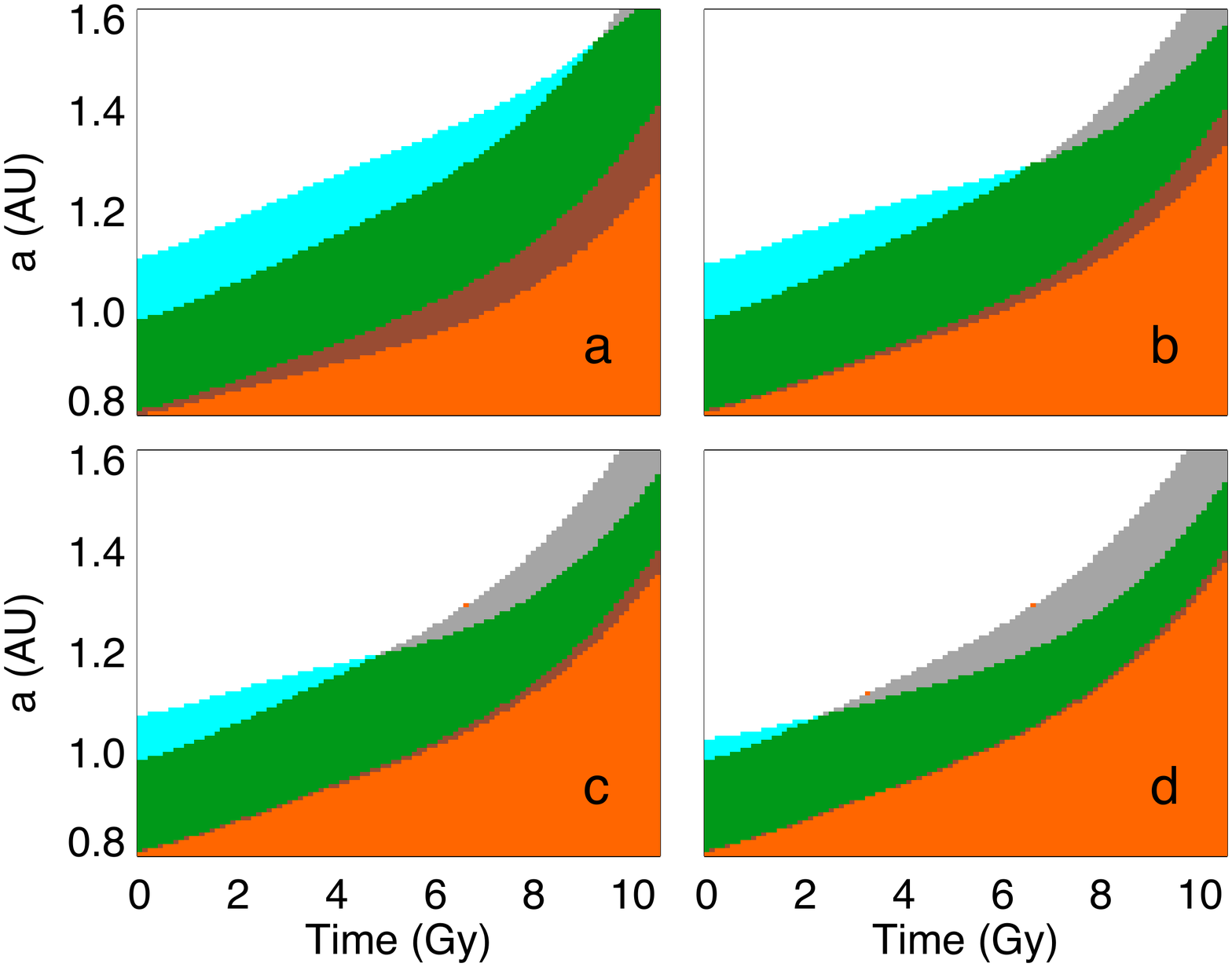}
\caption{Different climate states as a function of time and planetary distance $a$. The colors have the same meanings as in Figure~9.  The four panels show differences in the \co-pressure dependence of {\em land-based\/} weathering. Panels a, b, c, and d have $\beta_L=0.1$, 0.25, 0.35 and 0.5 respectively. In all cases, $W_{L0}=W_{S0}=1/6$. Values of other model parameters are the same as in Figure~9.}
\end{figure}

Figure 12 is similar to Figure~11 except that we examine changes in the \co-pressure dependence of {\em land-based\/} weathering rather than seafloor weathering. The panels show cases with $\beta_L=0.1$, 0.25, 0.35 and 0.5 respectively. In all cases, $\beta_S$ is kept fixed at a value of 0.3, and we use  $\theta_L=\theta_S=10$ K.

In contrast to Figure~11, the differences between the 4 panels in Figure~12 are relatively minor. In each case, a large fraction of phase space is occupied by the glaciated equilibrium state. The ice-free equilibrium state makes up a fairly narrow region that shrinks slightly at larger values of $\beta_L$. In panel a, the very weak land-based weathering \co-pressure dependence allows the inner edge of ice-free region to last longer. Towards the end of this phase, planets have very low \co\ levels, and can remain in ice-free equilibrium state with $P<10^{-5}$ bar for up to 1 Gy.

The main trends between the panels in Figure~12 are (i) growth in the size of the limit-cycle (grey) region; (ii) shrinking of the region in which both equilibria exist (cyan region); and (iii) a reduction in the fraction of the ice-free region that has \co\ pressures below $10^{-5}$ bar (brown region). The sizes of the ice-free and glaciated regions change only modestly as $\beta_L$ varies. This suggests that the \co-pressure dependence of land-base weathering plays a minor role in determining the climate when seafloor weathering is also present.

\begin{figure}
\plotone{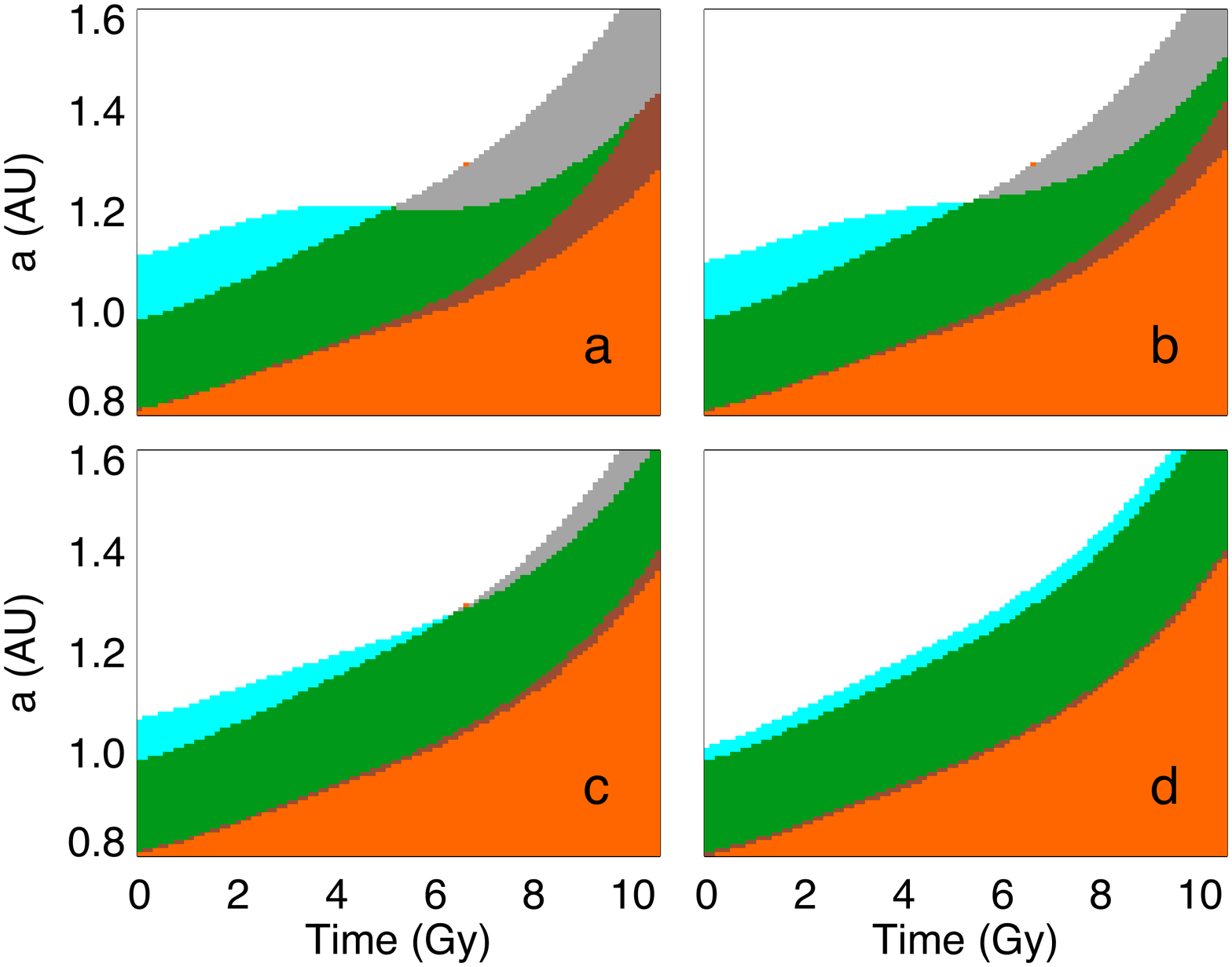}
\caption{Different climate states as a function of time and planetary distance $a$. The colors have the same meanings as in Figure~9. The four panels show differences in the initial outgassing rate $G_0$ compared to that on the modern Earth. Panels a, b, c and d use $G_0=30$, 10, 3, and 1 respectively. In all cases, $W_{L0}=W_{S0}=1/6$. Values of other model parameters are the same as in Figure~9.}
\end{figure}

Until this point, we have paid little attention to the initial magnitude of the outgassing rate. The outgassing rate is uncertain on the early Earth, and unknown for extrasolar planets. Figure~13 shows the effect of using different values for the initial outgassing rate $G_0$. Panels a, b, c and d use $G_0=30$, 10, 3 and 1 times the current rate on Earth. In the first 3 panels, the outgassing rate declines exponentially over time, matching the current rate on Earth at 4.5 Gy. The outgassing rate is constant in panel d.

Overall, the differences between the four panels of the figure are relatively minor, especially if the cyan region is considered as an extension of the white region in practice. For this set of model parameters, the outgassing rate plays a less significant role in determining the climate than other factors such as the seafloor-weathering \co-pressure dependence or the relative importance of land versus seafloor weathering.

The clearest difference between the panels is the decline in the wavy nature of the outer edge of the ice-free state (including the cyan region in which both equilibria exist). Moving through panels a to d, the variation in the outgassing rate over time is reduced, which has the same affect as increasing the relative importance of seafloor weathering that we saw in Figure~10. This is because seafloor weathering scales with outgassing, so increasing the contribution of seafloor weathering has the same result as making the outgassing rate less variable.

Note that the limit-cycle state occupies a relatively large fraction of phase space at late times in panel a of Figure~13. This occurs because the outgassing rate becomes very low late in the evolution in this case. Although the initial outgassing rate is very high, the outgassing rate is tied to that on Earth today at 4.5 Gy, so the decay timescale is short, and outgassing rates continue to decline quite rapidly at late times. As we saw earlier, limit cycles can occur when the outgassing rate is low and the insolation is not too large. 

For the same reason, the brown region, marking an ice-free equilibrium with very low $P$, is relatively prominent at late times in panel a. The outgassing rate is so low at this point that only a small amount of atmospheric \co\ is necessary for weathering to balance outgassing.

\begin{figure}
\plotone{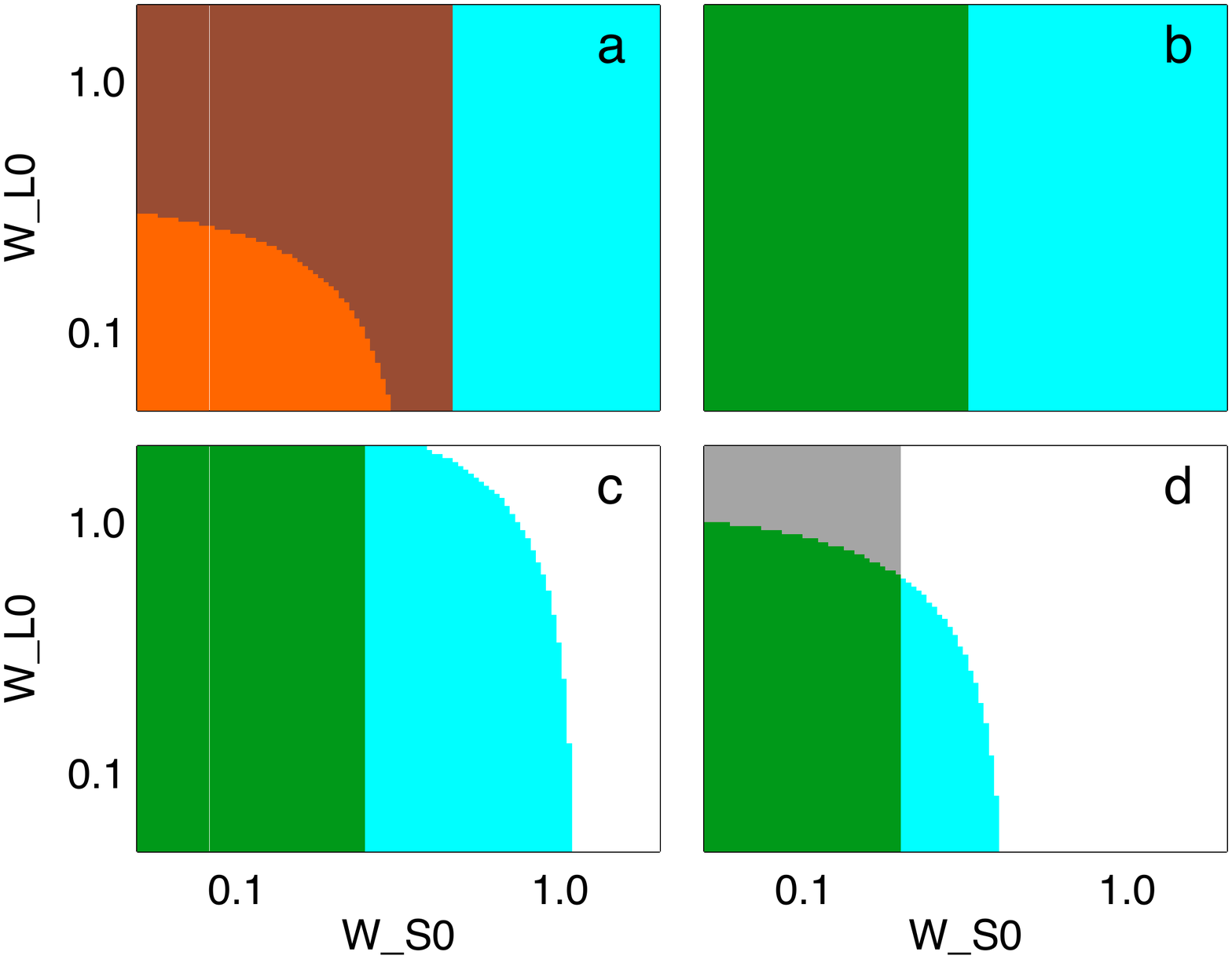}
\caption{Different climate states as a function of the intrinsic land and seafloor weathering rates $W_{L0}$ and $W_{S0}$. The colors have the same meanings as in Figure~9. The four panels show differences in the insolation level $S$ compared to that on the modern Earth. Panels a, b, c and d use $S=1.1$, 1, 0.9 and 0.8 respectively. In all cases, the outgassing rate is the same as the modern Earth.}
\end{figure}

%
%
\subsection{Dependence on Intrinsic Weathering Rate}
In the previous sections, we examined several cases involving different values of the intrinsic weathering rates $W_{L0}$ and $W_{S0}$, using values that illustrated interesting outcomes. We now briefly survey a wider range of possible intrinsic weathering rates since these quantities may vary substantially from one planet to another.

Figure 14 shows the different climate modes for a range of $W_{L0}$ and $W_{S0}$ values for four values of the insolation, ranging from $S/S_0=1.1$ in panel~a to $S/S_0=0.8$ in panel~d. In each case, the outgassing rate is the same as on Earth today. The climate states are color-coded using the same scheme as Figures 9--13.

Each panel of the figure can be divided into two regions depending on the rate of seafloor weathering. For seafloor weathering rates above a critical value, the ice-covered equilibrium state exists, while this state is absent for lower seafloor weathering rates. The ice-covered state may exist alone (white regions), or be accompanied by an equilibrium state that is ice-free (cyan regions). The minimum seafloor weathering rate necessary for the ice-covered state to exist decreases with decreasing insolation. Both equilibrium states exist concurrently over a wide range of phase space for panels a--c.

The total weathering rate (land plus seafloor) determines whether the ice-free equilibrium state can exist. If the insolation is high, and the total weathering rate is too low, then the climate will enter the runaway greenhouse state indicated by the orange region in panel~a. If the insolation is low, and the total weathering rate is too high, then the climate undergoes limit cycles between an ice-free and an ice-covered state, indicated by the grey region.

\begin{figure}
\plotone{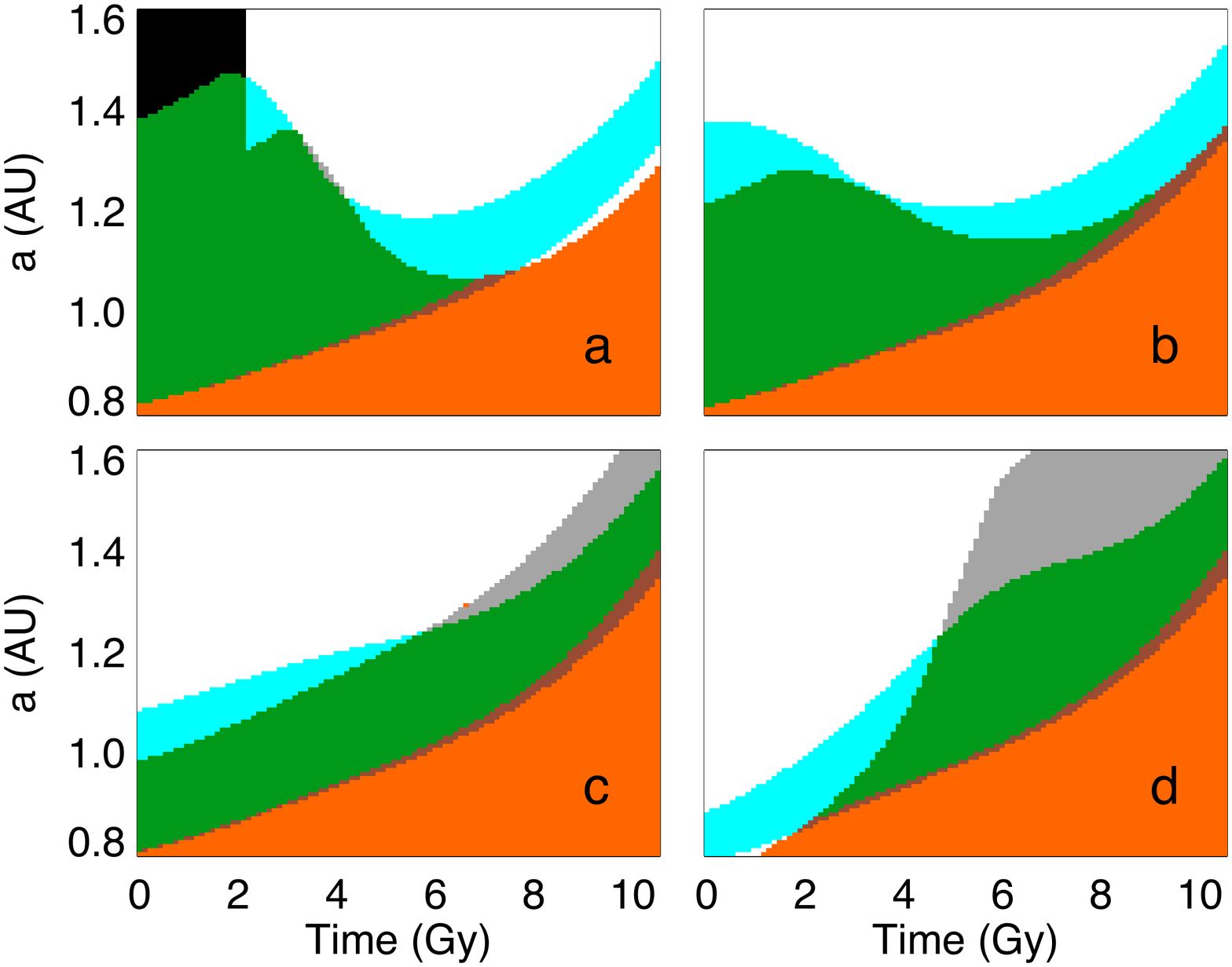}
\caption{Different climate states as a function of the relationship between seafloor weathering and outgassing. The colors have the same meanings as in Figure~9. In each panel, the seafloor weathering rate is proportional to $G^n$ where $G$ is the outgassing rate. Panels a, b, c and d use $n=0$, 0.5, 1 and 2 respectively. In all cases, the outgassing rate declines exponentially over time with $G_0=5$.}
\end{figure}

%
%
\subsection{Seafloor Weathering Relationship to Outgassing}
Until now, we have assumed that the seafloor weathering rate is proportional to the outgassing rate, since both are related to the seafloor spreading rate. Here, we examine other possible relationships.

Figure 15 shows examples in which the seafloor weathering rate is proportional to $G^n$, where $G$ is the outgassing rate. Panels a through d of the figure show cases with $n=0$, 0.5, 1 and 2 respectively. In each panel, the different climate states are color-coded using the same color scheme as Figures 9--14.

In panel a, seafloor weathering is independent of outgassing, so that it behaves in a similar manner to continental weathering. As a result, this case shares some similarities with panel a of Figure 10 which shows a case with continental weathering only. The extent of the ice-free state (green plus cyan plus brown regions) shows the same wave-like profile, covering a large range of orbital distances at early times, and a much narrower range at late times. The abrupt change at about 2 Gy is an artifact introduced by the inability of our climate model to include solutions with \co\ partial pressures above 10 bar.

The main difference between panel a of Figures 10 and 15 is that the latter includes an ice-free equilibrium state that extends over a large region of phase space. The region appears in all four panels of Figure~15, although the boundaries vary somewhat. Thus, the ice-covered equilibrium state is important for a wide range of possible relationships between seafloor weathering and outgassing.

Some changes are apparent as we increase the dependence of seafloor weathering on the outgassing rate. When weathering is independent of outgassing (panel a), the ice-free climate state is more apparent early in the planet's history. Weathering rates are relatively low at this point, so equilibrium \co\ levels are high enough to provide a habitable climate over a wide range of orbital distances. 

When the seafloor weathering rate depends strongly on outgassing (panel d), the ice-free equilibrium is more pronounced at later times. At early times, the high outgassing rate means that seafloor weathering is very rapid, allowing the glaciated state to exist for all the orbital distances considered here. At late times, the reduced outgassing rate means that seafloor weathering is very weak. The glaciated equilibrium state disappears as a result. Planets with moderate insolations exist in the ice-free state, while planets at large distances are prone to undergo limit cycles instead.

%
%
\subsection{Extent of the Glaciated Equilibrium State}
As we saw in the previous sections, much of the climate phase space of Earth-like planets is occupied by an equilibrium state in which the surface is covered in ice. This state exists when seafloor weathering is effective and the insolation is not too high. The glaciated state may be the only equilibrium state at a given time, or it can coexist with an ice-free equilibrium state. However, when both states exist, the glaciated state often appears first, so that the planet is likely remain ice covered after an ice-free solution appears, unless the climate receives a large perturbation.

Given the importance of the glaciated equilibrium we now derive an approximate analytic expression that gives the range of situations in which it is likely to occur. This expression may be used as a starting point for other studies of planetary climates when seafloor weathering is a factor.

The maximum insolation for which the glaciated state can exist is the insolation that generates a global temperature just below freezing for a surface albedo $\alpha_s$ appropriate for an ice covered surface. As before, we use $\alpha_s=0.605$. Tests using the climate model described in Section~2 show that the \co\ partial pressure of an ice-covered surface at 273 K is roughly related to the insolation $S$ by
\begin{equation}
\frac{S}{S_0}\simeq A\phi^3+B\phi^2+C\phi+D
\label{eq-sglaciated}
\end{equation}
where 
\begin{equation}
\phi=\ln\frac{P}{P_0}
\end{equation}
and $A=-3.8152\times 10^{-4}$, $B=-2.9847\times 10^{-3}$, $C=-2.8247\times 10^{-2}$, $D=1.2796$.

The only weathering that operates in this case is seafloor weathering. When seafloor weathering balances outgassing we have from Eqn.~\ref{eq-weathering}
\begin{equation}
G=W_{S0}G\exp\left[\beta_S\phi+\frac{\tsf-T_0}{\theta_S}\right]
\end{equation}
where
and $\tsf=283$ K and $T_0=288$ K. Thus at equilibrium we have
\begin{equation}
\phi=-\frac{1}{\beta_S}\left[\ln W_{S0}+\frac{\tsf-T_0}{\theta_S}\right]
\end{equation}

This expression for $\phi$ can be substituted into Eqn.~\ref{eq-sglaciated} to give the maximum insolation at which the glaciated equilibrium can exist.

\begin{figure}
\plotone{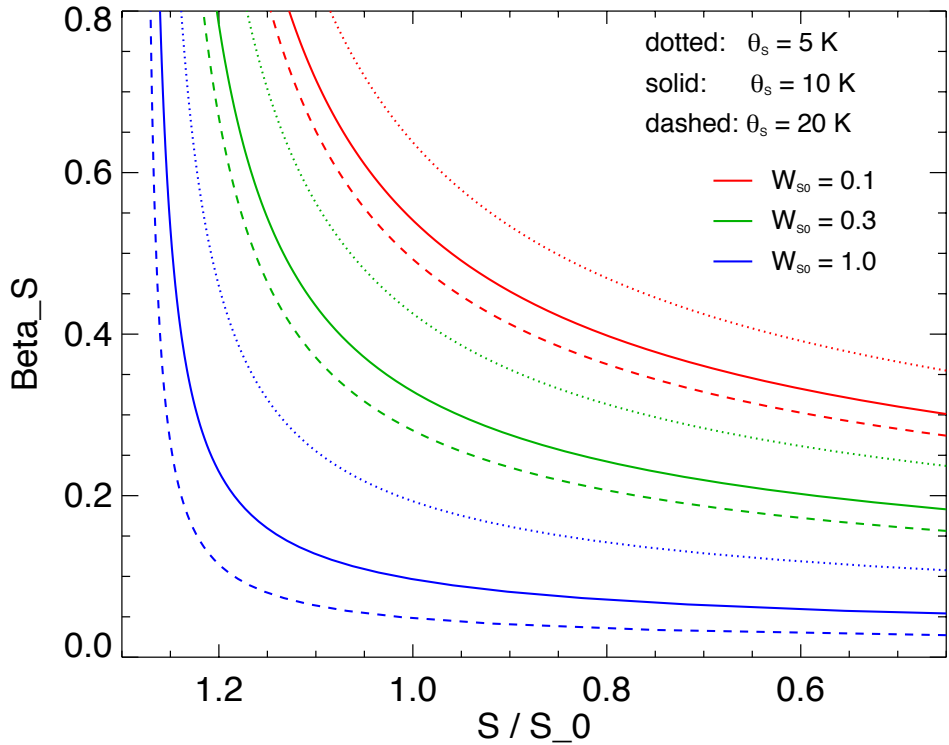}
\caption{Maximum insolation $S$ at which the ice-covered equilibrium state can exist for different values of the seafloor weathering parameters $W_{S0}$, $\beta_S$ and $\theta_S$. Regions above and to the right of each curve can exist in the glaciated state.}
\end{figure}

Figure 16 shows the maximum insolation for the glaciated equilibrium state for different seafloor weathering \co-pressure dependences $\beta_S$. Regions of phase space above and to the right of each curve can exist in the globally glaciated state, while regions to the left of the curve cannot. Note that the absence of a glaciated equilibrium doesn't guarantee that a stable ice-free climate exists---the climate could undergo limit cycles, but it will not be permanently covered in ice.

The solid curves in Figure 16 show cases with $\theta_S=10$ K, the dashed curves have $\theta_S=20$ K, and the dotted curves have $\theta_S=5$ K. Red, green and blue curves indicate cases with $W_{S0}=0.1$, 0.3 and 1 respectively. 

The maximum insolation for the glaciated equilibrium increases when the intrinsic rate of seafloor weathering is stronger (larger $W_{S0}$), as we would expect. When the intrinsic weathering rate is high, the \co\ pressure must be low in order to match the outgassing rate, leading to a relatively weak greenhouse effect. For all but small values of $\beta_S$, the glaciated state can persist for insolations 20\% higher than on Earth today when $W_{s0}=1$. Lower intrinsic weathering rates reduce the range of the glaciated state.

For large values of the weathering \co-pressure-dependence $\beta_S$, the extent of the glaciated region varies little with the value of $\beta_S$. The values of the other weathering parameters also make little difference in this case. Whether or not the ice-covered state exists depends mainly on the insolation level. However, when $\beta_S$ is small, a slight change in any of the weathering parameters can dramatically alter the extent of the ice-covered state. The climate will be highly sensitive to the precise details of seafloor weathering  in this situation.

%
%
\section{Discussion}

%
%
\subsection{Does Seafloor Weathering Occur on Ice-Covered Planets?}
In Section~3, we saw that one of the most frequent outcomes of the model is a climate characterized by an ice-covered surface where volcanic outgassing of \co\ is balanced by removal via seafloor weathering. Given the prevalence of this state, it is worth taking some time to examine whether seafloor weathering is likely to operate on an ice-covered world.

The first question is whether the ocean will freeze completely during a prolonged glaciation, or whether there is a layer of liquid water beneath the ice, allowing weathering reactions between sea water and the seafloor to continue. The surface of the ice may be far below freezing, but deeper layers will be warmer due to heat escaping from the planet's interior. We can estimate the thickness of the ice layer $\Delta z$ using a simple thermal conductivity calculation
\begin{equation}
\frac{\Delta T}{\Delta z}\simeq\frac{q}{K}
\end{equation}
where the thermal conductivity of ice is roughly $K\simeq2.3$ W/m/K. The current heat flux under the seafloor on Earth is about $q=0.1$ W/m$^2$ \citep{pollack:1993}. This gives an ice thickness of about 1.4 km for $\Delta T=60$ K, which is the maximum difference between the surface temperature and the freezing point seen in Figures~5--8. The ice layer is significantly thinner than the mean ocean depth of 3.7 km on Earth\footnote[1]{NOAA, http://ngdc.noaa.gov/mgg/global/etopo1\_ocean\_volumes.html}. We expect the ocean to remain partially liquid so that seafloor weathering continues to operate unless a planet's geothermal heat flow is much less than on Earth today, or the ocean is substantially shallower.

It is possible that varying ocean depth and ice-sheet dynamics mean that only some regions of the ocean avoid complete freezing, especially for a low geothermal heat flux. This may not matter in practice since seafloor weathering mainly occurs in regions that are relatively close to geothermally active regions \citep{coogan:2015, coogan:2018}, and these are the regions most likely to avoid total freezing due to their higher than average heat flux. These regions will also be near areas where outgassing occurs, providing a ready supply of both \co\ and weatherable rock.

It seems likely that seafloor weathering will continue for heat fluxes that are not too small. However, for seafloor weathering to regulate the climate there also needs to be effective exchange of \co\ between the atmosphere and the ocean. 

There are several possible ways that gas exchange between the atmosphere and ocean could be maintained. Some studies suggest that ice sheets may not extend all the way to the equator during glaciated epochs, leaving a belt of unfrozen ocean that would allow coupling with the atmosphere to continue \citep{abbot:2011}. The existence of large stretches of open water in the middle of the ocean may be unlikely since thick ice sheets at higher latitudes will flow into ice-free areas due to gravity \citep{hoffman:2017}. However, regions of open water could exist near land, depending on the geography of the coastline and proximity to geothermally active areas. \citet{lehir:2008} show that only a few thousand square km of open water are required for effective gas exchange between the atmosphere and oceans.

Even on a world with a completely ice-covered ocean, gas can diffuse through cracks in the ice, especially at fissures generated where grounded ice meets floating ice. Atmospheric \co\ can also dissolve in melt water and be transported into the ocean along continental edges \citep{hoffman:2017}. Melting may be especially prevalent in areas where windswept dust accumulates on the ice, darkening the surface, and raising the local temperature.

If the atmosphere and ocean fail to exchange \co\ effectively during a global glaciation it could lead to an interesting new dynamic. Seafloor weathering will continue in the ocean, moderating the level of dissolved \co\ in seawater. Outgassing on land will proceed without weathering, so \co\  in the atmosphere will increase to high levels.

Once the atmospheric \co\ level becomes high enough, the ice will melt, and oceans will once again be in contact with the atmosphere. Continental weathering will begin to reduce atmospheric \co\ levels. However, \co\ will also be absorbed into the ocean, which will have a low \co\ concentration due to earlier seafloor weathering. This drawdown of atmospheric \co\ into the ocean is likely to occur more rapidly than loss due to continental weathering.

The climate will return to the glaciated state once the atmospheric \co\ level is low enough. The end result is quite like the limit-cycle state caused by continental weathering in which the climate oscillates between an ice-covered and an ice-free state \citep{menou:2015}. However, the ice-free phase will be substantially shorter in this case, with a timescale set by the atmosphere-ocean gas exchange timescale rather than the continental weathering timescale. If, during one of these limit cycles, circumstances change so that the ice-free equilibrium state becomes stable then one would expect the climate to transition to this state at this point.

%
%
\subsection{Model Limitations and Uncertainties}
The climate evolution of an Earth-like planet depends a great deal on the details of continental and seafloor weathering. It is plausible that weathering depends on temperature and \co\ pressure in a manner similar to Eqn.~\ref{eq-weathering}, and most previous studies have assumed weathering behaves this way. Unfortunately, the values of the parameters in this equation are not known precisely.

The largest uncertainties are the intrinsic weathering rates $W_{L0}$ and $W_{S0}$, and the \co-pressure dependences $\beta_L$ and $\beta_S$. The temperature dependences $\theta_L$ and $\theta_S$, and the typical rock temperature $\tsf$ at which seafloor weathering takes place, are also somewhat uncertain. To date, most experimental and field studies of weathering have concentrated on weathering of rocks and soils on land. Fewer measurements have been made for seafloor weathering. This is unfortunate because the results in Section~3 suggest that climate evolution is especially sensitive to the strength of seafloor weathering.

Biological activity can affect climate in a number of ways. For weathering on land, it is often difficult to isolate the contribution of vegetation to the weathering rate \citep{schwartzman:1989, kump:2000}. This makes it harder to estimate weathering rates on planets that lack land plants. Weathering reactions on the seafloor depend on the chemical composition of seawater. This may have changed substantially on Earth within the last Gy due to the rise of sponges and other animals that remove dissolved silica. This probably altered the rate of reactions that compete with carbonate precipitation, so-called ``reverse weathering'', altering the effectiveness of seafloor weathering as a result \citep{isson:2018}. Changes in climate in turn affect the biosphere, which suggests that feedbacks may need to be considered \citep{caldeira:1992b, o'malley:2013}.

In some circumstances, continental weathering rates will be limited by the rate of supply of unweathered rock by erosion. This limitation may prevent the establishment of an equilibrium between outgassing and weathering if the outgassing rate is high or if the initial inventory of \co\ is large \citep{foley:2015}. We have not considered this possibility here, although the model does include supply limitations for seafloor weathering. On planets with substantial seafloor weathering, supply limitations on land-based weathering are probably not the main factor in determining the climate.

A final area of uncertainty is the temporal behavior of the planet itself. Specifically, how do the outgassing rate and the intrinsic weathering rates vary with time? In this study, we adopted an exponentially declining outgassing rate, which is consistent with simple models for mantle evolution \citep{kadoya:2015}. However, the exponential decay timescale and initial outgassing rate are uncertain and will differ from one planet to another. We also used fixed intrinsic weathering rates. However, weathering rates probably vary over time on tectonically active planets, and these variations have been linked to changes in climate behavior such as temporary global glaciations on Earth \citep{donnadieu:2004, mills:2011}.

All these uncertainties mean we cannot make confident predictions about the climate on a particular planet, even for the ancient Earth. However, we can establish the kinds of climate evolution that can take place, and what factors these depend on, and this is the main goal of this study. It also appears that some of the uncertain parameters have comparable effects on the climate, such as the weathering temperature and \co-pressure dependences described in Section~3.3. This reduces the complexity of the problem somewhat.

In addition to the uncertainties with regard to weathering, the models used here for the climate and the \co-cycle involve several  simplifications that may limit the scope of the results. For example, we model the climate using a single global surface temperature rather than temperature that varies spatially. This may be a reasonable approximation for determining whether the surface is ice-covered or ice-free because models that account for latitudinal variations tend to find that partially ice-covered states are unstable \citep{caldeira:1992a, spiegel:2008, haqq-misra:2016}. However, these studies focus on planets with low obliquities and rotation rates similar to Earth. Planets with high obliquities or different rotation rates, and thus different latitudinal heat transport rates, are likely to behave differently \citep{williams:1997, spiegel:2008, spiegel:2009}. The same is true for planets with atmospheric pressures very different than the 1 bar N$_2$ atmosphere assumed here \citep{vladilo:2013}.
The climate evolution of planets orbiting M stars may also be quite different than the cases considered here due to the different spectral energy distribution of these stars \citep{shields:2013}, and the possibility that planets in the habitable zone are tidally locked \citep{checlair:2017}.

In this study, we have assumed that the seafloor weathering rate is proportional to the outgassing rate based on the assumption that outgassing scales with the seafloor spreading rate  \citep{sleep:2001}. One result of this scaling is that while rapid, early outgassing makes the climate warmer if the only weathering is land-based, the same is not true when seafloor weathering is dominant because weathering rates are also high at early times. The actual relationship between outgassing and seafloor weathering may be more complicated than assumed here, and this will affect the climate as a result.

Another factor not considered here is the possibility that ocean sedimentation rates change during glaciated episodes. This will affect the rate of seafloor weathering. Sedimentation rates are probably reduced during glaciations due to the lack of transport into the ocean from rivers. This will be partially offset by glacial melting in dust-laden areas. 

Currently on Earth, only young seafloor weathers effectively since older seafloor is too deeply covered in sediment \citep{coogan:2015, coogan:2018}. A lower sedimentation rate may boost seafloor weathering rates by making more seafloor accessible. Countering this effect is the fact that reduced sedimentation reduces thermal blanketing of the seafloor, lowering the temperature of the rock exposed to seawater, and reducing the weathering rate. For very low sedimentation rates, the entire seafloor may be weatherable, and the weathering rate will be limited solely by the rate at which new seafloor forms, independent of \co\ pressure and temperature.

%
%
\subsection{Comparison with Previous Work}
\citet{kadoya:2015} examined the climate of a tectonically active planet as a function of time and orbital distance, including continental weathering but not seafloor weathering. They used a mantle cooling model to calculate the outgassing rate, which declines over time. The results of \citet{kadoya:2015} are similar to Figure 10a, which  shows a case with continental weathering but no seafloor weathering. The two models exhibit an ice-free equilibrium state, limit cycles, and a runaway greenhouse, occupying roughly similar regions of age-orbital distance phase space in each case. However, the behavior becomes more complex when seafloor weathering is included, as in Fig 10b-d. A new glaciated equilibrium state appears in some regions of phase space. In addition, the extent of the ice-free equilibrium state and the limit cycle mode depend strongly on the relative strengths of continental and seafloor weathering.

\citet{menou:2015} also found that limit cycles are common, and may occur for planets as close at 1.1 AU from a Sun-like star. Limit cycle periods range from tens to hundreds of My, with a planet spending most of the time in the frozen state. In common with the results of Section~3.3, \citet{menou:2015} found that the long-term evolution of ice-free climates depends on how the weathering rate varies with \co\ pressure. A strong \co-pressure dependence leads to large changes in temperature on Gy timescales, while a weak dependence means the equilibrium temperature is restricted to a narrow range, as we saw in Figure~7.

\citet{haqq-misra:2016} modeled climate evolution subject to land-based weathering using higher outgassing rates than assumed by \citet{menou:2015}. These authors also noted that the \co\ level in soil on Earth is increased by biological processes, so that continental weathering rates could be lower by a factor of about 5.5 on an abiotic Earth. As a result, they conclude that limit cycles are unlikely on abiotic planets with Earth-like outgassing rates. However, limit cycles could occur if the outgassing rate is lower or the continental weathering rate is higher. Figure 10a, which uses a somewhat higher intrinsic weathering rate (1/3 of the current value rather than 1/5.5), suggests that the limit-cycle state can occur for some planets undergoing land-based weathering when the outgassing rate is comparable to or lower than on Earth today.

\citet{haqq-misra:2016} showed that adding seafloor weathering for a planet undergoing limit cycles will increase the frequency of these cycles, but the authors did not explore the extent to which seafloor weathering could lead to a stable glaciated state.

\citet{abbot:2016} used a simple, linearized climate model to explore a wide range of weathering and outgassing rates and insolation values. He found that the extent and characteristics of a stable ice-free climate depend quite sensitively on the climate weathering parameters, and that these typically play a larger role in determining the climate than the outgassing rate. This agrees with the limited importance of the outgassing rate seen in Figure~13. \citet{abbot:2016} noted that a stable glaciated state could exist if seafloor weathering operates when the surface is frozen, although this possibility was not explored in detail.

\citet{kadoya:2019} examined the evolution of a planet undergoing continental weathering if the planet begins in a glaciated state rather than assuming an ice-free initial condition. Planets located far from the star are likely to begin in this ``cold start'' state because there is a maximum degree to which \co\ can warm a planet.  As the stellar insolation increases over time, a stable ice-free climate can appear. However, cold-start planets require a substantially higher insolation to transition to this ice-free state than a warm-start (ice-free) case because the albedo of an ice-covered planet is higher. Such a planet would be potentially habitable, but unable to enter the habitable state without a large climate perturbation.

In this paper, we found that seafloor weathering provides an additional barrier to accessing a stable ice-free climate. In this case, a cold-start planet can remain glaciated at higher insolations than considered by \citet{kadoya:2019} because seafloor weathering reduces the \co\ level. When seafloor weathering operates, a planet may never reach the maximum possible \co\ greenhouse. Even some planets that enjoy a warm start can become glaciated for billions of years if seafloor weathering is strong enough to counteract outgassing. 

%
%
\subsection{Implications}
Weathering on the seafloor has received relatively little attention compared to continental weathering in studies of planetary habitability. Does the addition of seafloor weathering change the big picture when it comes to habitability? The short answer is yes. The presence of seafloor weathering can lead to a new stable climate state in which a planet is completely or almost completely covered in ice. This equilibrium state does not exist in the absence of seafloor weathering because land-based weathering ceases during a global glaciation, and there is no process to balance outgassing.

Previous studies have noted that planets with low outgassing rates or high weathering rates may have no stable equilibrium climate \citep{mills:2011, menou:2015, abbot:2016}. Instead, the climate undergoes limit-cycles, oscillating back and forth between ice-free and ice-covered states. The glaciated equilibrium state caused by seafloor weathering overlaps much of the phase space that would be occupied by the limit-cycle regime in the absence of seafloor weathering. (This can be deduced from Figures~10 and 11.) Both of these climate regimes are uninhabitable according to the usual definition.

An important difference is that the ice-covered state tends to appear early in a planet's history when the insolation is low. Conversely, the limit-cycle state tends to exist at late times when the outgassing rate has declined due to cooling of the planet's interior. If a planet enters the glaciated equilibrium state at some point, it may not escape, even if an ice-free equilibrium state appears subsequently (e.g. the purple curves in Figure~8). An event, such as an impact, that melted the surface without appreciably altering the composition of the atmosphere would not alter the long-term situation because the surface will refreeze on a timescale much shorter than the time required to build up enough atmospheric \co\ to sustain an ice-free surface. Escaping from the ice-covered climate will require a prolonged period of increased outgassing, or a sustained decrease in the seafloor weathering rate, neither of which is guaranteed. Thus, some potentially habitable planets will be effectively uninhabitable because they are stuck in the wrong equilibrium.

We also note that because the glaciated climate state often occurs early in a planet's history, it could be detrimental to habitability if some key steps in the origin and evolution of life are more likely to occur when a planet is young.

Seafloor weathering is worth studying further since the existence of the glaciated equilibrium state introduces more diversity into the possible climate histories of a planet. Land-based and seafloor weathering also have different properties. One is correlated with the outgassing rate, while the other isn't. Continental weathering depends on the surface temperature, while seafloor weathering depends on the temperature of rocks beneath the seafloor.  This means each type of weathering needs to be considered and modeled separately. Climate evolution can also to be sensitive to the relative intrinsic strengths of land-based and seafloor weathering, even when the total intrinsic weathering rate is fixed (Figure~10).

There is good evidence that Earth has entered a global or near-global ice-covered state in its history \citep{hoffman:1998}. However, these ``snowball Earth'' episodes were short-lived. In contrast, planets in this study that enter the globally glaciated state typically remain there for billions of years. The reason for this difference is presumably that seafloor weathering was not strong enough to maintain the kind of ice-covered equilibrium that we saw in Section~3. Seafloor weathering would have increased the length of time it took for Earth's climate to leave the snowball state, but apparently seafloor weathering by itself was not sufficient to balance outgassing when the snowball episodes occurred. Instead, the climate briefly entered the limit-cycle state characterized by alternating glaciated and ice-free states. 

For the snowball Earth episodes that occurred within the last Gy, the climate apparently completed multiple limit cycles \citep{hoffmann:2004, mills:2011}, perhaps due to prolonged era of enhanced continental weathering or reduced outgassing. This deviation from the otherwise clement climate experienced over most of Earth's history suggests that intrinsic weathering and/or outgassing rates can change significantly on timescales of several hundred My, perhaps due to changing continental configurations. Comparable changes in the intrinsic seafloor weathering rate could force a planet into the glaciated equilibrium state, especially if the insolation is somewhat lower than on Earth today.

%
%
\section{Summary}
In this paper, we have examined the climate evolution of geologically active, Earth-like planets subject to continental and seafloor weathering. The main conclusions of this work are:
\begin{itemize}
\item A planet's climate is determined by a combination of insolation from the planet's star and the strength of the greenhouse effect due to \co\ and water vapor. Carbon dioxide in the atmosphere/ocean system is increased by outgassing, and decreased by weathering. Temperature-dependent weathering acts to stabilize the climate against changes in insolation and the outgassing rate.

\item Continental weathering probably ceases when the surface temperature falls below freezing. However, seafloor weathering should continue especially around ice-free oases. Here, geothermal heating maintains liquid water in contact with the seafloor, and allows \co\ exchange between the ocean and atmosphere.

\item In our model, two kinds of equilibrium climate exist. One state is characterized by an ice-free surface, with outgassing balanced by  land-based and seafloor weathering. The other state consists of an ice-covered surface where outgassing is balanced by seafloor weathering only. The second of these has not been studied in detail before.

\item Depending on the insolation, outgassing rate, and weathering parameters, one or both equilibrium states can exist for the same planet. If the insolation is too high, the climate will enter a runaway greenhouse instead.

\item Stellar luminosity increases over time, so a planet typically encounters the ice-covered equilibrium state before the ice-free state. In the absence of large perturbations, a planet will remain in the ice-covered state as long as it exists, even if the ice-free state also appears subsequently. Many potentially habitable planets may not be habitable in practice for this reason.

\item In some regions of phase space, neither equilibrium is stable, and the climate cycles between ice-covered and ice-free states. This  typically happens when the outgassing rate is low or the intrinsic weathering rates are high.

\item The climate behavior depends strongly on the \co-pressure dependence of seafloor weathering, which is poorly constrained. The climate is also sensitive to the relative contributions of continental and seafloor weathering even when the total intrinsic weathering rate is held constant.
\end{itemize}

%
%
\acknowledgments

I would like to thank Lindsey Chambers and an anonymous reviewer for comments and discussions that have greatly improved this paper.

%
%

\end{document}